\documentclass[fleqn,usenatbib]{mnras}

\usepackage{newtxtext,newtxmath}

\usepackage[T1]{fontenc}
\usepackage{ae,aecompl}
\usepackage{hyperref}

\usepackage{graphicx}	
\usepackage{amsmath}	
\usepackage{amssymb}	

\newcommand{\vsini}{\ensuremath{{\upsilon}\sin i}}
\newcommand{\kms}{\,km\,s$^{-1}$}

\begin{document}

\title{A revisit to the enigmatic variable star 21 Comae}

\author[Paunzen et al.]{Ernst Paunzen,$^{1}$
Gerald Handler,$^{2}$
Przemys{\l}aw Walczak,$^{3}$
Stefan H{\"u}mmerich,$^{4,5}$
\newauthor
Ewa Niemczura,$^{3}$
Thomas Kallinger,$^{6}$
Werner Weiss,$^{6}$
Klaus Bernhard,$^{4,5}$
\newauthor
Miroslav Fedurco,$^{7}$
Anna G{\"u}tl-Wallner,$^{6}$
Jaymie Matthews,$^{8}$
Theodor Pribulla,$^{9}$
\newauthor
Martin Va\v{n}ko,$^{9}$
Stefan Wallner,$^{6,10}$
Tomasz R{\'o}{\.z}a{\'n}ski$^{3}$
\\
$^{1}$Department of Theoretical Physics and Astrophysics, Masaryk University, Kotl\'a\v{r}sk\'a 2, 611 37 Brno, Czech Republic\\
$^{2}$Nicolaus Copernicus Astronomical Center, Bartycka 18, 00-716 Warsaw, Poland\\
$^{3}$Instytut Astronomiczny, Uniwersytet Wroc{\l}awski, 51-622 Wroc{\l}aw, Poland\\
$^{4}$American Association of Variable Star Observers (AAVSO), Cambridge, USA\\
$^{5}$Bundesdeutsche Arbeitsgemeinschaft f{\"u}r Ver{\"a}nderliche Sterne e.V. (BAV), Berlin, Germany\\
$^{6}$Institute of Astronomy, University of Vienna, T{\"u}rkenschanzstrasse 17, 1180 Vienna, Austria\\
$^{7}$Faculty of Science, P. J. {\v S}af{\'a}rik University, Park Angelinum 9, Ko{\v s}ice 040\,01, Slovak Republic\\
$^{8}$Department of Physics and Astronomy, University of British Columbia, 6224 Agricultural Road, Vancouver, BC V6T 1Z1, Canada\\
$^{9}$Astronomical Institute, Slovak Academy of Sciences, 059 60 Tatransk{\'a} Lomnica, Slovak Republic\\
$^{10}$ICA, Slovak Academy of Sciences, Dubravska cesta 9, 84\,503 Bratislava, Slovak Republic
}

\date{Accepted XXX. Received YYY; in original form ZZZ}

\pubyear{2019}

\label{firstpage}
\pagerange{\pageref{firstpage}--\pageref{lastpage}}
\maketitle

\begin{abstract}
The magnetic chemically peculiar (Ap/CP2) star 21\,Com has been extensively studied in the past, albeit with widely differing and sometimes contradictory results, in particular concerning the occurrence of short term variability between about 5 to 90 minutes. We have performed a new investigation of 21\,Com using MOST satellite and high-cadence ground-based photometry, time series spectroscopy, and evolutionary and pulsational modeling. Our analysis confirms that 21\,Com is a classical CP2 star showing increased abundances of, in particular, Cr and Sr. From spectroscopic analysis, we have derived $T_{\rm eff} = 8\,900\pm200$\,K, $\log g = 3.9\pm0.2$, and $\vsini = 63\pm2$ km\,s$^{-1}$. Our modeling efforts suggest that 21\,Com is a main sequence (MS) star seen equator-on with a mass of $2.29\pm0.10$\,M$_{\odot}$ and a radius of $R$\,=\,$2.6\pm0.2$\,\rm{R}$_{\odot}$. Our extensive photometric data confirm the existence of rotational light variability with a period of 2.05219(2)\,d. However, no significant frequencies with a semi-amplitude exceeding 0.2\,mmag were found in the frequency range from 5 to 399\,d$^{-1}$. Our RV data also do not indicate short-term variability. We calculated pulsational models assuming different metallicities and ages, which do not predict the occurrence of unstable modes. The star 18\,Com, often employed as comparison star for 21\,Com in the past, has been identified as a periodic variable ($P$\,=\,1.41645\,d). While it is impossible to assess whether 21\,Com has exhibited short-term variability in the past, the new observational data and several issues/inconsistencies identified in previous studies strongly suggest that 21\,Com is neither a $\delta$ Scuti nor a roAp pulsator but a ``well-behaved'' CP2 star exhibiting its trademark rotational variability.
\end{abstract}


\section{Introduction}

Several studies have been concerned with the occurrence of pulsational variability in chemically peculiar (CP) stars in the past. In the non-magnetic Am/Fm (CP1) stars, $\gamma$ Doradus and $\delta$ Scuti pulsations are regularly observed \citep[e.g.][]{2011A&A...535A...3S,Balona2011,2013MNRAS.429..119P,2017MNRAS.465.2662S,2017MNRAS.466.1399H}. The magnetic Bp/Ap (CP2) stars, on the other hand, are known to exhibit rapid oscillations (high-overtone, low-degree, and non-radial pulsation modes) in the period range of about 5-20 min. Stars exhibiting this kind of variability are referred to as rapidly oscillating Ap (roAp) stars \citep{1982MNRAS.200..807K}. However, there are only very few studies that report the occurrence of $\gamma$ Doradus and $\delta$ Scuti pulsations in CP2 stars, and some of these results remain controversial, as discussed below.

\citet{Kurtz2008} claimed the first observational evidence for $\delta$ Scuti pulsations and a magnetic field in the CP2 star HD\,21190. However, \citet{Bagnulo2012} considered both the magnetic field detection and the CP2 classification of this star spurious. Recently, \citet{Niemczura2017} showed that this star is in fact a physical binary system for which the complex interplay between stellar pulsations, magnetic fields and chemical composition has to be taken into account. Using $Kepler$ data, \citet{2011MNRAS.410..517B} reported the presence of $\gamma$ Doradus and $\delta$ Scuti pulsations in several CP2 stars. However, the authors caution that the CP2 nature of these stars needs to be confirmed by new spectroscopic observations. \citet{Neiner2015} reported a 76(13)\,G magnetic field for HD 188774, a $\delta$ Scuti/$\gamma$ Doradus hybrid pulsator \citep{Lampens2013}. This star is a very interesting object because its spectral type of A7.5\,IV-III clearly places it beyond the terminal-age MS, which is confirmed by the very precise Gaia parallax \citep{Gaia16,Gaia18,Luri18} available. HD 41641 has been identified as a $\delta$ Scuti star with chemical signatures of a mild CP2 star \citep{Escorza2016}. However, until now, no magnetic field measurements are available for this object. Finally, \citet{Neiner2017} reported the clear detection of a magnetic field with a longitudinal field strength below 1\,G for the evolved Fm (CP1) $\delta$ Scuti pulsator $\rho$ Pup (HD 67523). This star has often been studied in the past, with several contradicting results (see \citealt{Yushchenko2015} for an overview).

The atmospheres of CP2 stars are enriched in elements such as Si, Cr, Sr or Eu, and anomalies are usually present as surface abundance patches or spots \citep{Deutsch58,Pyper69,Goncharskii83}, probably connected with the globally organized magnetic field geometry \citep{Michaud1981}.\footnote{We note that not in all cases the surface structures are connected to the magnetic field topology \citep{Kochukhov19}} This leads to photometric variability with the rotation period (e.g. \citealt{Krticka2013}), which is explained in terms of the oblique rotator model \citep{Stibbs1950} and referred to traditionally as $\alpha^{2}$ Canum Venaticorum (ACV) type variability.

In this paper, we present a new and detailed photometric and spectroscopic study of the enigmatic CP2 star 21 Com, which has been extensively studied since 1953, albeit with widely differing and sometimes contradictory results. 21\,Com has been found to exhibit ACV variability; however, published rotational periods range from 0.9 to 11\,d. Furthermore, several studies claimed the existence of short-term variability with periods typical of $\delta$ Scuti and roAp star pulsations. Nevertheless, the existence of pulsational variability in 21\,Com has remained a controversial issue and widely differing claims as to the underlying period(s) have been published (see Section \ref{history}).

If the published claims hold true, 21\,Com would be an outstanding object from an astrophysical point of view because it shows ACV, $\delta$ Scuti and roAp-type variability - a combination not known for any other star. To investigate the literature claims in more detail and with modern instrumentation, a wealth of new ground and space based photometric as well as spectroscopic data were acquired. Photometric variability was investigated on time scales from minutes to days. Furthermore, we analyzed radial velocity measurements and performed an elemental abundance analysis based on high-resolution spectra. State-of-the-art pulsational models were computed and compared with the observational results.

An overview of the literature results for our target star is presented in Section \ref{history}. Data sources and frequency analysis are described in Section \ref{sources}. Abundances analysis, astrophysical parameters, pulsational models and a discussion on the evolutionary status are presented in Section \ref{Models}. We conclude in Section \ref{Conclusion}.

\begin{table*}
\begin{center}
\caption{The results of former investigations on 21\,Com and the present study. The dashes (``--'') in columns five to seven indicate that the corresponding frequency range was not investigated.}
\label{data_history}
\begin{tabular}{ccccccccc}
\hline
\hline
References & Phot & Filter & Spec & ACV & $\delta$ Scuti & roAp & Amplitude & Comp \\
& & & & [d] & [min] & [min] & [mmag] & \\
\hline
\citet{1953ApJ...118..489P} & & & * & 7.75 & -- & -- &  & none \\
\citet{1955PASP...67..342D} & & & * & 1.0256 & -- & -- & & none \\
\citet{1957ZA.....41..254B} & * & $V$ & & 11, 2.1953, 1.10 & 32 & -- & 10 & 18\,Com \\
\citet{1972AJ.....77..666B} & * & $UBV$ & & 2.1953 & -- & -- & 20 -- 40 & 18\,Com, 22\,Com \\
\citet{1973AA....22..381P} & * & $BV$ & & -- & 30 & -- & 20 & 18\,Com \\
\citet{1975AJ.....80..698P} & * & $BV$ & & -- & 30.4, 39.5 & -- & 10 -- 15 & 18\,Com \\
\citet{1976SvA....19..734A} & & & * & 1.03 & -- & -- & & none \\
\citet{1980AA....90...18W} & * & $UBV$ & & 0.9178 & 32.7 & -- & 7 -- 21 & 18\,Com \\
\citet{1981AN....302..219T} & * &  & & -- & 22 & -- &  & none \\
\citet{1982CoKon..83..190J} & * &  & & -- & constant & -- &  & none \\
\citet{1982IBVS.2237....1M} & * & $B$ & & -- & constant & 5.4, 5.9 & 1.5 & 17\,Com B, 18\,Com \\
\citet{1983IBVS.2368....1G} & * & $U$ & & -- & 24 & 6 & 4.5 & none\\
\citet{1983AA...128..152W} & * &  & & -- & constant & -- &  &  \\
\citet{1987CoSka..16....7Z} & * & $\beta$ & & 1.837 & 90 & -- & 19 & 18\,Com \\
\citet{1989IBVS.3373....1S} & * & none & & -- & 36 & constant &  & none \\
\citet{1990MNRAS.245..642K} & * & $UBVbv$ & & 2.00435 & constant & constant & $\le$\,30 & 18\,Com \\
\citet{1993AJ....105.1903N} & * & $B$ & & -- & -- & constant & & none \\
\citet{1993MNRAS.263..742V} & * & $B$ & & -- & 37, 65, 95 & constant &  $\le$\,4.5 & TYC\,1989-1807-1 \\
\hline
This work & * & MOST,$v$ & * & 2.05219 & constant & constant & 19 & 18\,Com, 22\,Com \\ 
\hline
\hline
\end{tabular}
\end{center}
\end{table*}

\section{21\,Com and its history} \label{history}

\citet{Cannon1920} first identified 21\,Com (HD\,108945, HR\,4766, UU\,Com, $V=5.44$\,mag) as a CP2 star (spectral type A3pSr), which has been confirmed by many subsequent studies. The star is currently listed in the Catalogue of Ap, HgMn and Am stars with a spectral type of A3pSrCr \citep{RM09}.

21\,Com has first been identified as a physical member of the open star cluster Melotte 111 in Coma Berenices by \citet{Trumpler38}. Based on Hipparcos/Tycho data, \citet{Kharchenko04} estimated the membership probability as 93\,\% from proper motions and 100\,\% from the location of the star in the colour-magnitude diagram and in the sky. The star's membership to Melotte 111 has later also been confirmed by \citet{Silaj2014}. Melotte 111 is a very close (distance between 85 and 100\,pc from the Sun) open cluster that has a very controversial age determination. Literature values range from 400 to 800\,Myr \citep{Tang2018}, which is astonishing but in line with the estimates for other open clusters in the solar vicinity \citep{Netopil2015}.

\citet{Shorlin2002} reported the detection of a very weak magnetic field (109(44)\,G) in 21\,Com. This was confirmed later on by \citet{Landstreet2008a} on the basis of a clear non-zero signal in Stokes $V$. 21\,Com is thus an excellent target to study the minimum magnetic field strength needed to develop surface stellar spots detectable via photometry.

Since 1953, almost twenty publications have dealt with the variability of 21\,Com using photometric and spectroscopic measurements. For a discussion of these results, it is important to distinguish between the aims of the different studies. Observations were typically optimized to either search for/analyse the long-term (ACV) or the short-term ($\delta$ Scuti and/or roAp) variations. Variations on both time scales were studied in only a few investigations.

We have investigated all individual references and detected several inconsistencies with already published overviews \citep[e.g.][]{1990MNRAS.245..642K}. For convenience, variability ranges were divided into an ACV ($P$\,$>$\,0.5\,d), $\delta$ Scuti (10\,min$<$\,$P$\,$<$\,0.5\,d) and roAp ($P$\,$<$\,10\,min) region. Table \ref{data_history} summarizes the literature results for 21\,Com and is organized as follows:
\begin{itemize}
\item Column 1: reference.
\item Column 2: an asterisk indicates the use of photometric data.
\item Column 3: employed filters.
\item Column 4: an asterisk indicates the use of spectroscopic data.
\item Column 5: period [d] of the rotationally-induced (ACV) variability.
\item Column 6: period [min] of the $\delta$ Scuti variability.
\item Column 7: period [min] of the roAp variability.
\item Column 8: amplitude [mmag].
\item Column 9: employed comparison star(s).
\end{itemize}

\begin{figure}
\begin{center}
\includegraphics[width=0.47\textwidth]{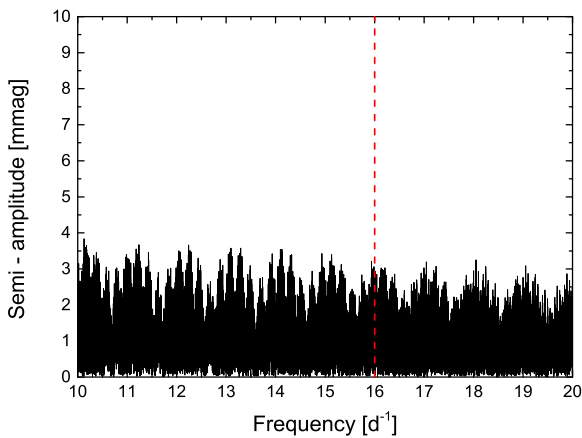}
\caption{Frequency spectrum of the \citet{1987CoSka..16....7Z} data that have been prewhitened by the correct rotational period of 2.05219 d (cf. Eq. \ref{eq1}). No significant signals in the period range around 90 mins ($\sim$16\,d$^{-1}$; indicated by the dashed line) remain.}
\label{zverko}
\end{center}
\end{figure}

It has to be emphasized that most studies were concerned with the investigation of a certain frequency domain, for example the search for $\delta$ Scuti pulsation \citep{1983AA...128..152W}. All photometric investigations were carried out with a photomultiplier (i.e. photoelectrically), which has been extensively and successfully employed for light variability studies of CP stars. Attained measurement accuracies were generally of the order of a few mmags \citep{Blanco1978,Dukes2018}. In general, as regards methodology, the past studies on 21\,Com have been very heterogeneous, e.g. in regard to the employed instrumentation, observing cadence, or the (non-)use of comparison stars. The binarity study of \citet{Abt1999}, which -- to our knowledge -- is the only work that concluded that 21\,Com is a spectroscopic binary (SB) system, is discussed in more detail in Section \ref{Abt_binarity}.

In summary, the ACV variability of 21\,Com has been well established because of its large amplitude (about 0.02\,mag). Most published periods are either approximately 1.1 or 2.2\,d. As ACV variables are prone to exhibiting double-waved light curves in several wavelength regions \citep{Maitzen1978,Leone2001}, this discrepancy can be readily explained. 

The situation is much less clear for the period range between a few to 90 minutes, and contradictory results have been obtained over almost 40 years. Most strikingly, on the same level of photometric accuracy and with the same set of filters, some studies reported variability on diverse time scales while others reported constancy. This unsatisfactory situation has probably constituted the main cause why 21\,Com has not been observed during the past 25 years. In the following, we exemplarily discuss the results of some studies devoted to the short period ($\delta$ Scuti and roAp) region.

Due to the authors listing only very few details, an assessment of the very short periods reported by \citet{1982IBVS.2237....1M} is very difficult. The authors analyzed more than 450 measurements obtained on 60 nights, but neither data, light curves nor frequency spectra were presented, and the lack of information on the observing cadence renders it impossible to check whether the derived periods are related to the sampling frequency and thus represent aliases of the rotational frequency of the star. We note, however, that the authors caution that, due to the low S/N ratio of the variability signal, further independent proof of the reality of their results is necessary.

From an analysis of photometric observations taken during eight nights in 1978 and 1979, \citet{1987CoSka..16....7Z} reported a rotational period of 1.83736 d. He furthermore confirmed the existence of short term variations and, after prewhitening his data with the (incorrect) rotational period, derived possible short periods around 90 min. Using the original data from Table 3 of \citet{1987CoSka..16....7Z}, we reanalyzed the data employing the correct rotational period of 2.05219 d (cf. Eq. \ref{eq1}). Interestingly, no significant signals in the corresponding period range around 90 mins remain (Fig. \ref{zverko}). We would like to emphasize that \citet{1987CoSka..16....7Z} discussed \textit{possible} periods and stressed the preliminarity of his analysis -- an important piece of information that has unfortunately got lost in later review papers \citep[e.g.][]{1990MNRAS.245..642K}.

\begin{figure}
\begin{center}
\includegraphics[width=0.47\textwidth]{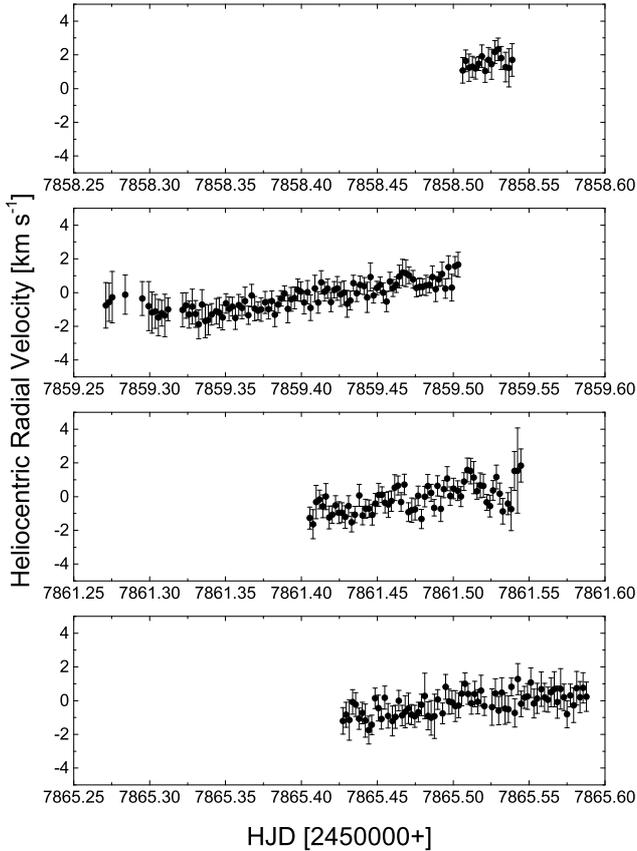}
\caption{Measured radial velocities of 21\,Com as a function of time.}
\label{RV_curve}
\end{center}
\end{figure}

\citet{1993MNRAS.263..742V} analyzed observations made on five nights in 1987 and six nights in 1988 and concluded that the star exhibits short-term variability with periods of 37 min, 65 min, and 95 min. Unfortunately, the authors did not list their observations in tabular form so that it was not possible to duplicate their results and we had to rely on the presented light curves. Only the light curves of data taken during four nights are illustrated, and there seems to be some confusion as to the observational date: the same dataset is identified with two different dates in the figures (February 10, 1988 versus February 9, 1988; compare the upper panels of Figs. 2 and 8 in \citealt{1993MNRAS.263..742V}). Except for some rather sudden excursions, the depicted light curve bands in Figs. 2 and 8, which illustrate the nightly observations taking during time intervals of about two to four hours, look rather constant and exhibit non-negligible scatter. Only from visual inspection, the presence of $\delta$ Scuti-type variations in these data is, of course, hard to assess. However, bearing in mind the short observational time span, the simultaneous fitting of the nightly data with three frequencies appears a rather optimistic approach and the authors caution that further investigation is required in regard to the presence of $\delta$ Scuti variability in 21\,Com.

In summary, as these examples illustrate, there seems to be no hard evidence for the presence of short period signals in the very heterogeneous body of work that has been concerned with the variability of 21\,Com. Some of the main concerns we could identify were (a) time-series analysis based on very short data sets, (b) the use of no comparison/check stars, and (c) the use of a variable comparison star (18\,Com; cf. Section \ref{18Com}). In some cases at least, these issues seem to have led to overinterpretation of the available data.


\section{Data sources and frequency analysis} \label{sources}

The following sections give an overview over the data sources and the employed methods of frequency analysis. Here, and throughout the paper, frequencies are given to the last significant digit.


\subsection{Radial velocity observations}

Radial velocity (RV) measurements were obtained using the {\'e}chelle eShel spectrograph of the Star{\'a} Lesn{\'a} Observatory (High Tatras, Slovakia), which is part of the Astronomical Institute of the Slovak Academy of Sciences. The spectrograph is attached to a 60 cm (f/12.5) Zeiss reflecting telescope (G1 pavilion). The employed CCD camera 
(ATIK 460EX) uses a 2750x2200 chip resulting in a resolving power between 11\,000 and 12\,000 within a spectral range 4\,150 to 7\,600 \AA. The reduction of the raw frames and extraction of the 1D spectra using the IRAF package tasks, Linux shell scripts, and FORTRAN programs, have been described by \citet{Pribulla2015}. The wavelength reference system, as defined by the preceding and following Th-Ar exposures, was stable to within 0.1\,$\mathrm{km\,s}^{-1}$.

RVs were measured with our own implementation of the two-spectra cross-correlation function (CCF) technique TODCOR \citep{Zucker1994}, for which a synthetic spectrum was calculated with the atmospheric parameters and abundances given in Sect. \ref{abundance_analysis}. We crosschecked the derived RV values with those of the FXCOR task within IRAF, which yielded essentially identical results. The resulting RV curves are shown in Figure \ref{RV_curve}.

\subsection{MOST observations} \label{data_MOST}

The Microvariablity and Oscillations of Stars (MOST) mission is a Canadian microsatellite designed to detect stellar oscillations with micromagnitude precision in bright stars. The spacecraft is equipped with a 15-cm Rumak-Maksutov telescope that feeds two Marconi 47-20 frame-transfer CCD devices (1024x1024 pixels). Observations are taken through a custom broadband filter with a wavelength coverage from about 3\,500 to 7\,000\,\AA. MOST is in a Sun-synchronous polar orbit, which allows the monitoring of stars between --19$^\circ$ and +36$^\circ$ declination for up to two months. It has an orbital period of about 101\,min. Being dedicated entirely to asteroseismology, MOST has contributed (and still is contributing) significantly to our understanding of stellar atmospheres and the internal composition of stars. For more information on the MOST satellite, the reader is referred to \citet{2003PASP..115.1023W}.

\begin{figure}
\begin{center}
\includegraphics[width=0.47\textwidth]{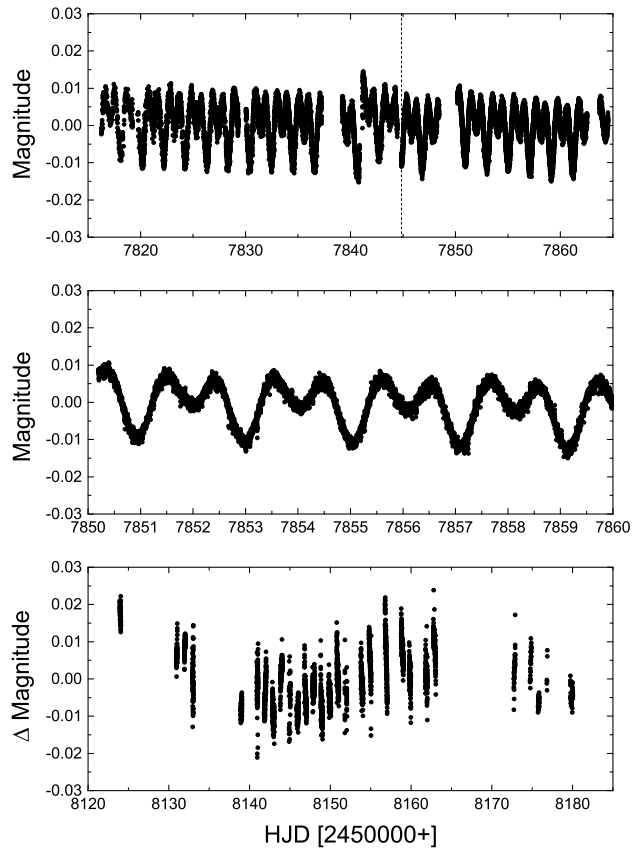}
\caption{Light curves of 21\,Com. The panels illustrate, from top to bottom, the full MOST light curve, a zoomed-in view of a 10-day segment of the MOST light curve, and the full APT light curve, respectively. Obvious outliers were removed from MOST data by visual inspection. The dashed line in the upper panel indicates HJD 2457844.8635, from which date onwards 21\,Com was observed exclusively and the cadence consequently increased to near-continuous coverage. The long-term trend seen in the APT measurements is due to beating of the 2-d rotation period and the nightly sampling of the data.}
\label{lightcurve}
\end{center}
\end{figure}

21\,Com was observed in Direct Imaging (DI) mode for a timespan of 48.27 days (HJD 2457816.2560--2457864.5274), during which a total number of 31\,816 observations were procured. Prior to HJD 2457844.8635 (indicated by the dashed line in the upper panel of Fig. \ref{lightcurve}), the observing time was split between 21\,Com and another star, which resulted in a reduced duty cycle and conspicuous data spacing in the corresponding part of the data set. From HJD 2457844.8635 onwards, 21\,Com was observed exclusively and the observing cadence consequently increased to near-continuous coverage.

While orbiting the partially illuminated Earth, MOST is subjected to variable amounts of stray light. This leads to artificial signals at multiples of the orbital frequency \citep{Reegen2006}. Long-term mean magnitude shifts (``jumps'') are also sometimes present, which were also apparent in the data for 21\,Com. We have tried to augment this situation by decorrelating against instrumental parameters, such as temperature. While this reduced the amplitude of the intrumental trends, they are still apparent in the data. The MOST light curve of 21\,Com is illustrated in Figure \ref{lightcurve} (upper and middle panels). The final light curve has a relative point-to-point scatter of $\sim$0.75\,ppm.

\subsection{Ground-based photometry} \label{gb_phot}

We acquired photoelectric time-series observations of 21\,Com with the 0.75-m Automatic Photoelectric Telescope (APT) T6 at Fairborn Observatory in Arizona. Differential photometry was collected through the Str{\"o}mgren $v$ filter, with 18\,Com (C1, HD\,108722, $V=5.47$\,mag, F5\,IV) and 22\,Com (C2, HD\,109307, $V=6.24$\,mag, A4\,Vm) as comparison stars. Because we wished to sample both the long-term rotational modulation of 21\,Com as well as possible roAp oscillations properly, an observing sequence C1 - Sky - V - C2 - V - C1 - Sky - V and so on, with two integrations of 20\,s each for each star, was chosen. That way the first data sampling alias occurs at 798 c/d, i.e. an effective Nyquist frequency of 399\,d$^{-1}$ (period of 3.6 min) is reached. As the measurement of sky background destroys the equidistant sampling of the variable star's measurements, Nyquist aliases are very low in amplitude ($<$ 30\% of an intrinsic peak). Therefore, these data are well suitable for frequency searches above the nominal Nyquist frequency.

The data were reduced in a standard way, starting by correcting for coincidence losses, sky background and extinction. A standard extinction coefficient of 0.34 mag/air mass was used. Then, differential light curves were created and the timings were corrected to the heliocentric frame. It turned out that 18\,Com is itself a variable (see Sect. \ref{18Com}). Consequently, this variability was removed from the 21\,Com data before proceeding with the analysis. The final data set for 21\,Com comprised 3\,969 differential measurements with an rms error of 1.9 mmag per single data point, and spanned 59 days from January to March 2018 (31 nights of actual observation).


\subsection{18\,Com - a new variable star} \label{18Com}

\begin{figure}
\begin{center}
\includegraphics[width=0.47\textwidth]{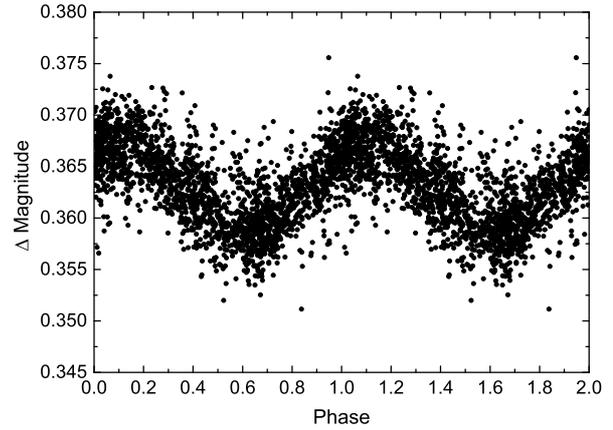}
\caption{Photometric measurements of 18\,Com differential to 22\,Com as a function of phase. The data have been  detrended and folded with the mean frequency of 0.70599\,d$^{-1}$ ($P$\,=\,1.41645\,d).}
\label{18Com_phase_plot}
\end{center}
\end{figure}

As mentioned above, we here report, for the first time, the variability of 18\,Com (HR\,4753), which has often been employed as comparison star for measuring 21\,Com during several decades (Sect. \ref{history}). Our photometric measurements are indicative of variability with a frequency of 0.70599\,d$^{-1}$ ($P$\,=\,1.41645\,d) and an amplitude of 8\,mmag (Fig. \ref{18Com_phase_plot}), in addition to a slower trend in the light curve. 

Given the period and light curve shape, there are three possibilities that may explain the variability of 18\,Com: pulsation, rotational variation, or ellipsoidal variability with an orbital period twice the photometric period. For their examination some knowledge of the basic stellar parameters is required; several determinations are given in the literature. In chronological order, \citet{Reiners06} derived $T_{\rm eff} = 6490$\,K, $M_v=1.71$\,mag from Str\"omgren photometry and $v \sin i = 97 \pm 5$\,\kms, hence $P_{rot} / \sin i = 1.74$\,d spectroscopically. \citet{Robinson07} give $T_{\rm eff} = 6535$\,K, log $g = 3.68$, and $[Fe/H]=+0.32$ from low-resolution spectroscopy. \citet{Casagrande11} derived $T_{\rm eff} = 6676$\,K, log $g = 3.68$, and $[Fe/H]=+0.25$ from colour-metallicity-temperature and colour-metallicity-flux calibrations. Meanwhile, an accurate Gaia DR2 parallax ($16.72\pm0.10$ mas; \citealt{Gaia16,Gaia18,Luri18}) became available that leads to $M_v=1.54\pm0.03$. All these determinations are in good agreement with each other. We can therefore estimate the stellar radius with $3.44\pm0.11$\,R$_{\odot}$ propagating the errors above. The evolutionary tracks used by \citet{Casagrande11} imply a mass of $1.93\pm0.05$\,M$_{\odot}$ for 18 Com.

Turning to the possibility of an ellipsoidal variable, we make two ``extreme'' assumptions for the secondary star, a 0.25\,M$_{\odot}$ M dwarf and a 1.0\,M$_{\odot}$ early G-type star. With an orbital period of 2.833\,d, Kepler's Third Law yields orbital semimajor axes of 10.9 and 12.1 \,R$_{\odot}$ for the two cases, respectively. Therefore, the primary's orbital velocity will be 22.4/73.5\,\kms, respectively. A frequency analysis of the radial velocities of 18\,Com listed by \citet{Abt1976} yields no periodic variability in excess of 9.3\,\kms. Furthermore, for our two scenarios, Equation 6 of \citet{Morris1985} indicates a photometric semi-amplitude of ellipsoidal variations of 5.8 and 17.0 mmag/sin$^2$\,i, respectively. In summary, if 18\,Com had a companion in a 2.833-d orbit massive enough to generate ellipsoidal variations with the measured amplitude, the data by \citet{Abt1976} should have revealed its presence in radial velocity, which, however, is not the case.

Investigating the case for rotational variability, the answer is already hidden within the basic parameter determinations quoted above. A rotational period of 1.416\,d in our data is consistent with the constraint $P_{rot} / \sin i = 1.74$\,d and yields an inclination of the rotation axis of $\approx 54^o$. As regards to pulsation, the pulsation constant Q for a 1.416\,d variation would be 0.31\,d for 18\,Com, which is well within the range of known $\gamma$ Doradus-type pulsators (cf. Fig. 9 of \citealt{Handler02}). On the other hand, in the HR Diagram 18\,Com is located at the low-temperature/high-luminosity corner of the $\gamma$ Doradus instability strip \citep{Balona18gaia}, which would make it somewhat unusual.

We conclude that the variability of 18\,Com is best compatible with rotational variation; however, gravity-mode pulsation cannot be excluded on the basis of the available data. To separate the two hypotheses, time-resolved spectroscopy would be very valuable as it would allow to investigate whether we are dealing with nonradial pulsational or rotational variation. Time-resolved colour photometry, on the other hand, would allow to check whether any colour changes consistent with temperature variations caused by pulsation are present.

\subsection{21 Com - frequency analysis} \label{section_periodanalysis}

Unless indicated otherwise, we have used the software package \textsc{Period04}, which employs discrete Fourier transformation and allows least-squares fitting of multiple frequencies to the data \citep{period04}. To extract all relevant frequencies, the data were searched for periodic signals and repeatedly prewhitened by the most significant frequency. Periodogram features and corresponding phase plots were carefully examined to prevent instrumental signals from contaminating our results.

To our knowledge, no search for RV or line profile variations caused by $\delta$ Scuti type pulsation has been performed for 21\,Com (Sect. \ref{history}). An analysis of this kind of variations, which occur on time scales of up to two hours, has to be performed with care because the surface spots also create line profile variations \citep{Oksala2015}. However, the latter kind of variations are connected with the rotational period of about two days and hence occur on a longer time scale.

The search for periodicities in the line profile variations was performed with a generalized form of the least-squares power spectrum technique as described by \citet{Mantegazza1999}. We emphasize that the resolving power of our spectra (R $\approx$ 12\,000) is at the lower limit required for such an investigation, which is generally carried out on spectra of higher resolution \citep{Mantegazza2002}. Using several strong Fe and Si lines, we have found no variation exceeding 0.5\% from the mean combined spectrum of all observations.

\begin{figure}
\begin{center}
\includegraphics[width=0.47\textwidth]{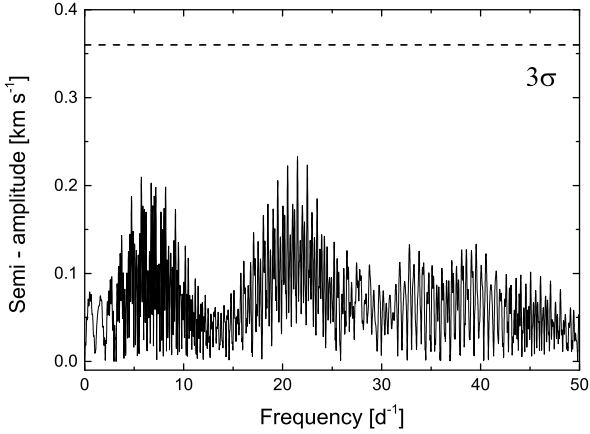}
\caption{Frequency analysis of the RV data for 21\,Com. Also indicated is the 3$\sigma$ level calculated from
the noise. The broad peaks at around 7 and 22\,d$^{-1}$ are only on a 1.8$\sigma$ level.}
\label{RV_Fourier}
\end{center}
\end{figure}

The RV data, on the other hand, are indicative of variability in the corresponding frequency range. As can be seen from Fig. \ref{RV_curve}, however, variations are only seen in small parts of the data. Furthermore, there is an obvious long-term trend, which might be related to either instrumental effects or the longer rotational period of 21\,Com. After removing the long-term trend from the data, the amplitude spectrum shown in Fig. \ref{RV_Fourier} has been derived. No statistically significant signals are present; the broad peaks at around 7 and 22\,d$^{-1}$ are only on a 1.8$\sigma$ level. Furthermore, even the higher frequency corresponds to a period of about 70 minutes, which is more than twice as long as the period(s) reported in the past (Table \ref{data_history}). In summary, no evidence for $\delta$ Scuti pulsation is present in our spectroscopic data. We caution, however, that this finding is to be viewed in the context of the limitations of the employed dataset.

\begin{figure}
\begin{center}
\includegraphics[width=0.47\textwidth]{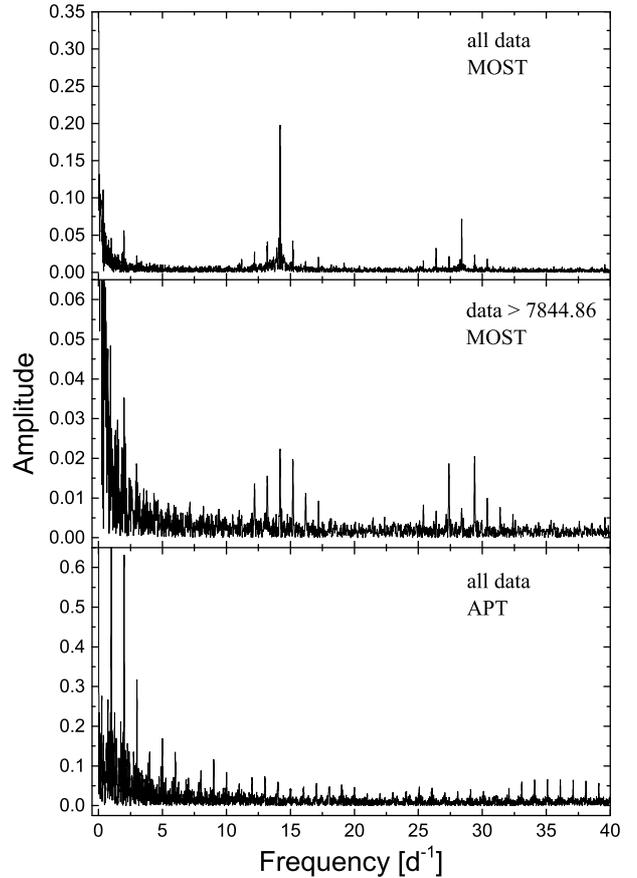}
\caption{Spectral windows of the complete MOST data set (upper panel), of MOST data obtained after HJD 2457844.8635 (middle panel) and the complete APT data set (lower panel). Note the different scales on the ordinates.}
\label{pa1}
\end{center}
\end{figure}

Figure \ref{pa1} illustrates the spectral windows of the complete MOST data set (upper panel), of MOST data obtained after HJD 2457844.8635 (middle panel) and the complete APT data set (lower panel). The spectral windows of the MOST data show a complex aliasing pattern that is dominated by peaks at the satellite's orbital frequency $f_{\rm{orb}}$\,=\,14.2\,d$^{-1}$ and its integer multiples, and surrounding alias peaks, which are separated from $f_{\rm{orb}}$ by 1, 2, and 3\,d$^{-1}$. Because of the near-continuous coverage achieved after HJD 2457844.8635, the orbital frequency signal is much reduced in the more recent data (Fig. \ref{pa1}, middle panel). As expected for single-site ground-based data, the spectral window for APT data is dominated by peaks at 1\,d$^{-1}$ plus corresponding integer multiples.

\begin{figure}
\begin{center}
\includegraphics[width=0.47\textwidth]{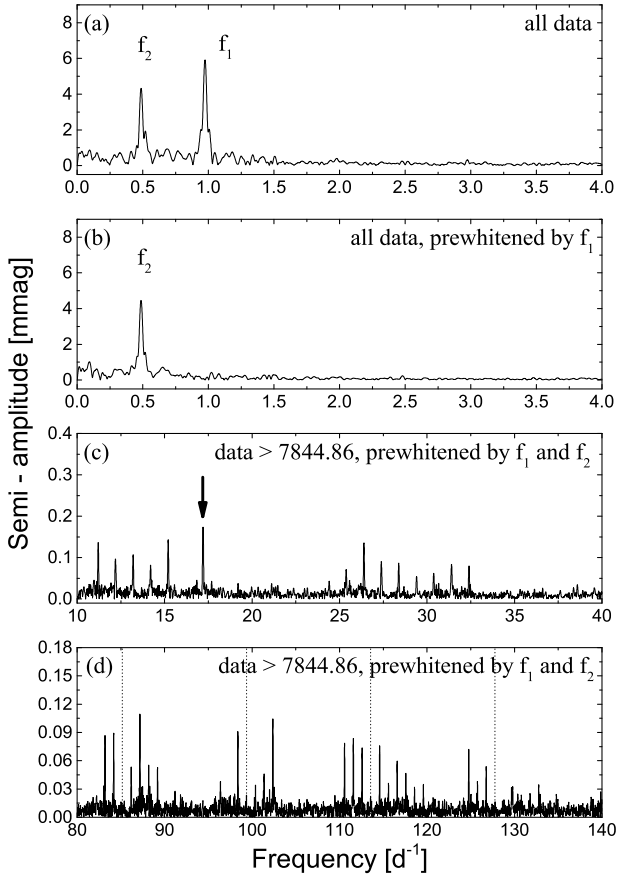}
\caption{Frequency analysis of MOST data for 21\,Com, illustrating important steps of the frequency spectrum analysis. The panels show, respectively, the frequency spectra of (a) unwhitened data, (b) data that has been prewhitened by $f_1$, (c) data obtained after HJD 2457844.8635 that has been prewhitened by $f_1$ and $f_2$, and (d) the same as in (c) but for a different frequency range. The vertical dotted lines in panel (d) denote integer multiples of the MOST orbital frequency. Note the different scales on the ordinates and the abscissae in the bottom panels. The ordinate axes denote semi-amplitudes as derived with \textsc{Period04}.}
\label{pa2}
\end{center}
\end{figure}

As first step, a frequency search was carried out in the range of 0\,<\,$f($d$^{-1})$\,<\,40 using the whole MOST dataset. After the identification of the first two dominant frequencies, $f_1$\,=\,0.97458(2)\,d$^{-1}$ and $f_2$\,=\,0.48729(3)\,d$^{-1}$ (Fig. \ref{pa2}, panels (a) and (b)), the search range was restricted to 0.2\,<\,$f($d$^{-1})$\,<\,40. Only $f_{\rm{orb}}$, its integer multiples and the corresponding 1\,d$^{-1}$ alias peaks could be extracted subsequently from the complete data set. 

As has been pointed out, the orbital frequency signal is much reduced in data obtained after HJD 2457844.8635. Therefore, we have restricted our search for small-amplitude signals to the more recent data. After prewhitening for the dominant frequencies $f_1$ and $f_2$, we have investigated the frequency range of 10\,<\,$f($d$^{-1})$\,<\,40. Again, no signal was found, only the above-mentioned multiples and alias peaks of $f_{\rm{orb}}$, albeit with significantly reduced amplitude (Fig. \ref{pa2}, panel (c)). The peak with the next highest amplitude in the investigated range is at 17.197443\,d$^{-1}$, which nearly exactly corresponds to an alias of $f_{\rm{orb}}$ (the corresponding peak is identified with an arrow in panel (c)). After that, only further frequencies related to $f_{\rm{orb}}$ can be identified. 

In addition, we have searched for the presence of high-frequency signals in the roAp star frequency domain (about 70 to 300\,d$^{-1}$) in the more recent MOST data taken after 
HJD 2457844.8635. The investigated range is dominated by signals at $f_{\rm{orb}}$, its integer multiples and the corresponding 1\,d$^{-1}$ alias peaks. Panel (d) of Fig. 
\ref{pa2}, which shows a sample part of the investigated frequency range, illustrates this situation. No significant frequencies could be extracted. The results of our frequency analysis of the MOST data are presented in Table \ref{pa_table}.

\begin{figure}
\begin{center}
\includegraphics[width=0.47\textwidth]{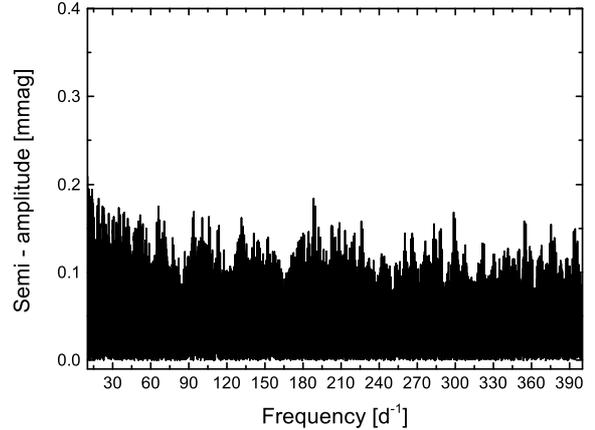}
\caption{Frequency spectrum of APT data for 21\,Com. APT data have been prewhitened by $f_1$ and $f_2$. The ordinate axes denote semi-amplitudes as derived with \textsc{Period04}.}
\label{pa3}
\end{center}
\end{figure}

APT data clearly confirm the two dominant low frequencies identified in the MOST dataset; within the errors, the same frequencies were derived. Furthermore, these data are especially suited to the identification of high-frequency signals and allow to expand the investigated frequency range up to 399\,d$^{-1}$. Accordingly, APT data were employed to search for high-frequency signals. No significant frequencies with a semi-amplitude exceeding 0.2\,mmag were found in the range from 5 to 399\,d$^{-1}$ (cf. Fig. \ref{pa3}). We interpret this as additional evidence that the frequency at 17.197443\,d$^{-1}$ in MOST data is related to the orbital frequency signal of the satellite, as expected.

In summary, we conclude that the photometric data for 21\,Com are indicative of two significant frequencies at $f_1$\,=\,0.97458(2)\,d$^{-1}$ and $f_2$\,=\,0.48729(3)\,d$^{-1}$, which are related to the star's rotational frequency. No other significant frequencies are present. The following elements were derived using the APT and high-cadence MOST data.
\begin{equation} \label{eq1}
\begin{aligned}
HJD (Min) = 2457852.9936(1) + 2.05219(2) \times E  
\end{aligned}
\end{equation}

\begin{table}
\begin{center}
\caption{Significant frequencies, semi-amplitudes and signal-to-noise ratios detected in 21\,Com, as derived with \textsc{Period04} from MOST data.}
\label{pa_table}
\begin{tabular}{cllll}
\hline
\hline
ID & Frequency  & Semi-amp. & S/N & Remark\\
   & [d$^{-1}$] & [mmag] & & \\       
\hline
$f_1$	& 0.97458(2) & 6.0 & 87.4 & $2f_{\mathrm{rot}}$ \\
$f_2$ & 0.48729(3) & 4.5 & 56.0 & $f_{\mathrm{rot}}$ \\
\hline
\hline
\end{tabular}
\end{center}
\end{table}

\subsection{On the possible binarity of 21\,Com} \label{Abt_binarity}

To our knowledge, the only work that identified 21\,Com as an SB system was published by \citet{Abt1999}, who found that the spectra indicate a double-lined binary with components always blended by moderate rotation (51 and 60\,km\,s$^{-1}$). Combined orbital elements were derived, which, however, are not very accurate. We emphasize that no eclipses were ever observed in 21\,Com.

\begin{figure}
\begin{center}
\includegraphics[width=85mm, clip]{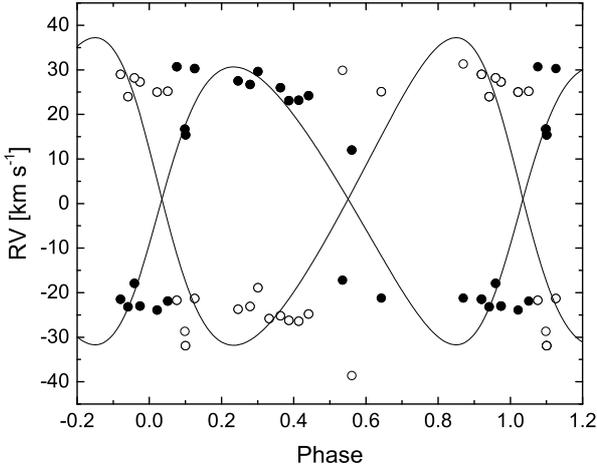}
\caption{Phase plot of the RV observations of \citet{Abt1999}, folded with the orbital parameters given by the same authors. Filled and open circles denote measurements for the proposed primary and secondary components, respectively.}
\label{orbit_data} 
\end{center} 
\end{figure}

In Fig.\,\ref{orbit_data}, we present a phase plot of the RV observations of \citet{Abt1999}, folded with the orbital parameters given by the same authors. The proposed orbital period is 18.813(10)\,d$^{-1}$ and thus very different from the rotational period. It is obvious that the behaviour of the RVs is not consistent with an SB system. Sharp, step-like transitions of both ``components'' occur at around phases $\sim$0.1 and $\sim$0.6, apart from which the RVs remain fairly constant. Some data points are out of the previously described trend. In these cases, however, the data points obtained at the same observational dates are influenced in the same way, which may indicate systematic errors.

For both components, the mean values of the RVs at the individual observational dates are constant, which results in a mean system velocity of +1.0(1.8)\,km\,s$^{-1}$. In summary, we conclude that the proposed SB nature of 21\,Com has probably been based on a misinterpretation of the line profile variations due to the stellar surface spots \citep{Oksala2015}. This is in line with the results from our spectroscopic analysis (cf. Section \ref{abundance_analysis}), which also do not indicate that the star is a double-lined spectroscopic binary.

\begin{figure*}
	\includegraphics[width=150mm]{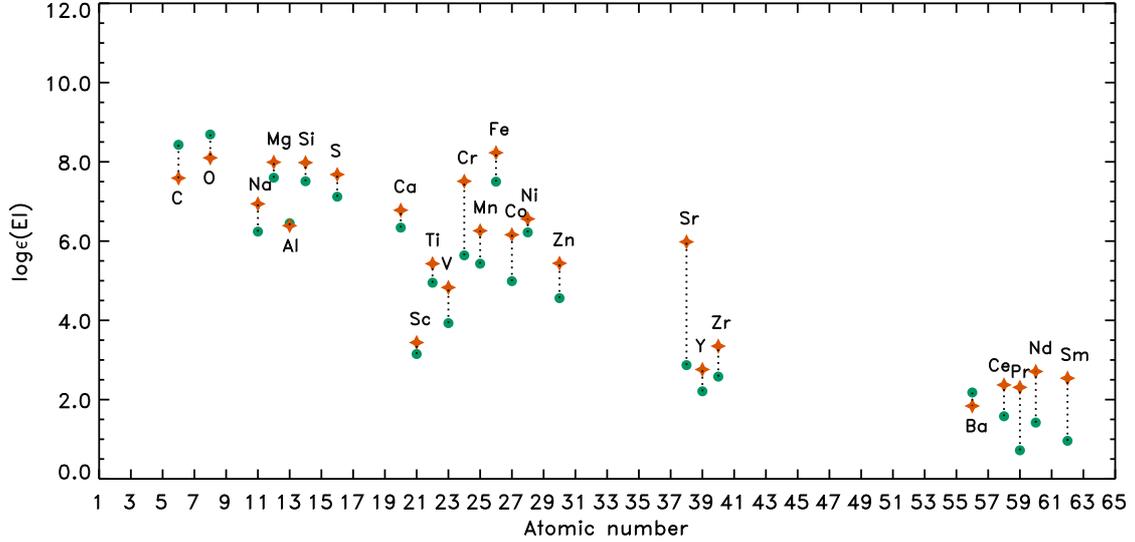}
	\caption{Comparison of the chemical composition of 21\,Com (orange stars) to the solar abundance pattern from \citet{AGSS09} (green circles). Overabundant and underabundant elements in 21\,Com are indicated, respectively, above and below their corresponding symbols.}
	\label{fig:abundances}
\end{figure*}

\begin{table}
	\begin{center}
	\caption{Chemical abundances and standard deviations for individual elements. Solar abundances were taken from \citet{AGSS09}.}
	\label{tab:abundances}
	\begin{tabular}{lcccc}
		\hline
		\hline
		Element & Z & No. of     & Abundance$^1$ & Solar \\
		        &   & lines used &               & abundance \\ 
						\hline	
C  &  6 &  2 &  $7.59 \pm 0.00$ & 8.43 \\
O  &  8 &  2 &  $8.10 \pm 0.00$ & 8.69 \\
Na & 11 &  2 &  $6.94 \pm 0.00$ & 6.24 \\
Mg & 12 &  8 &  $7.99 \pm 0.20$ & 7.60 \\
Al & 13 &  2 &  $6.39 \pm 0.00$ & 6.45 \\
Si & 14 &  9 &  $7.98 \pm 0.18$ & 7.51 \\
S  & 16 &  3 &  $7.68 \pm 0.09$ & 7.12 \\
Ca & 20 & 11 &  $6.78 \pm 0.18$ & 6.34 \\
Sc & 21 &  2 &  $3.44 \pm 0.00$ & 3.15 \\
Ti & 22 & 24 &  $5.43 \pm 0.18$ & 4.95 \\
V  & 23 &  1 &  $4.83 \pm 0.00$ & 3.93 \\
Cr & 24 & 52 &  $7.51 \pm 0.11$ & 5.64 \\
Mn & 25 &  5 &  $6.26 \pm 0.21$ & 5.43 \\
Fe & 26 & 58 &  $8.23 \pm 0.12$ & 7.50 \\
Co & 27 &  1 &  $6.16 \pm 0.00$ & 4.99 \\
Ni & 28 & 13 &  $6.56 \pm 0.31$ & 6.22 \\
Zn & 30 &  1 &  $5.44 \pm 0.00$ & 4.56 \\
Sr & 38 &  5 &  $5.98 \pm 0.15$ & 2.87 \\
Y  & 39 &  7 &  $2.76 \pm 0.24$ & 2.21 \\
Zr & 40 &  3 &  $3.35 \pm 0.48$ & 2.58 \\
Ba & 56 &  2 &  $1.84 \pm 0.00$ & 2.18 \\
Ce & 58 &  2 &  $2.37 \pm 0.00$ & 1.58 \\
Pr & 59 &  1 &  $2.31 \pm 0.00$ & 0.72 \\
Nd & 60 &  8 &  $2.71 \pm 0.20$ & 1.42 \\
Sm & 62 &  2 &  $2.54 \pm 0.00$ & 0.96 \\
		\hline
		\hline
\multicolumn{4}{l}{$^1$ given in $\log (N_{\mathrm{EL}}/N_{\mathrm{H}})+12.0$} \\
	\end{tabular}
\end{center}
\end{table}

\section{Abundance analysis, astrophysical parameters, evolutionary and pulsational models} \label{Models}

The following sections provide information on the chemical composition of our target star, its astrophysical parameters and the results from evolutionary and pulsational modeling as well as the employed methods.

\subsection{Abundance analysis and astrophysical parameters} \label{abundance_analysis}

The high resolving power (R\,=\,65\,000) spectra for the abundance analysis were obtained with ESPaDOnS (Echelle SpectroPolarimetric Device for Observations of Stars) in the spectral domain from 3\,700\,\AA\ to 10\,000\,\AA. Optical characteristics of the spectrograph and instrument performance are described by \citet{Donati2006}. The pipelined-reduced spectra were taken from the CFHT Science Archive\footnote{http://www.cadc-ccda.hia-iha.nrc-cnrc.gc.ca/en/cfht/}. All available spectra were averaged to minimize the effect of line profile variations due to stellar spots. Investigating both individual and mean spectra, we found no indications that the star is a double-lined spectroscopic binary (see also Section \ref{Abt_binarity}).

Effective temperature $T_{\rm eff}$ and surface gravity $\log g$ were determined from comparison of the observed and synthetic hydrogen $H\gamma$, $H\beta$ and $H\alpha$ lines \citep{Catanzaro2010}. To estimate the uncertainties of $T_{\rm eff}$ and $\log g$, we took into account the differences in the values obtained from separate Balmer lines, which result from validity of normalization.

Atmospheric parameters obtained from the hydrogen Balmer lines fitting were checked through an analysis of the Fe\,\textsc{I} and Fe\,\textsc{II} lines. In this step, effective temperature, surface gravity and microturbulence $\xi$ were changed until no difference between the iron abundances as determined from the different lines remained. It was not possible to use the trend of iron abundance versus excitation potential and the ionization equilibrium of Fe\,\textsc{I} and Fe\,\textsc{II} because of the relatively high projected rotational velocity \vsini\ of the star, which results in strong blending of spectral lines. We used the spectrum synthesis method, which allows for simultaneous determination of various parameters affecting the shape of the spectral lines, like $T_{\rm eff}$, $\log g$, $\xi_t$, \vsini, and chemical abundances. Atmospheric parameters have been set before the chemical abundance determination.

From the analysis of the hydrogen and iron lines, we have determined $T_{\rm eff} = 8\,900\pm200$\,K, $\log g = 3.9\pm0.2$, $\xi = 0.7\pm0.2$\,km\,s$^{-1}$ and $\vsini = 63\pm2$ km\,s$^{-1}$. The obtained chemical abundances compared with the solar abundances from \citet{AGSS09} are given in Table \ref{tab:abundances} and shown in Figure \ref{fig:abundances}. Our results fit in well with the parameters from former investigations (Table \ref{parameters_history}).

\begin{table}
\begin{center}
\caption{Astrophysical parameters of 21\,Com from former modern investigations (published during the past 20 years) and the present study. We note that several studies do not provide error estimates.}
\label{parameters_history}
\scriptsize
\begin{tabular}{cccc}
\hline
\hline
References & $T_{\rm eff}$ & $\log g$ & $\log L/$L$_{\odot}$ \\
           & [K]           &          &                      \\
\hline
\citet{Gebran2016} & 8\,300 & 3.3 & \\
\citet{Silaj2014} & 8\,800 & & 1.602 \\
\citet{Koleva2012} & 8\,906(1\,668) & 4.19(1.54) & \\
\citet{McDonald2012} & 8\,327 & & 1.53 \\
\citet{Lipski2008} & 8\,750 & & \\
\citet{Netopil2008} & 8\,700(240) & & \\
\citet{Kochukhov2006} & 8\,950(300) & & 1.71(7) \\
\citet{Adelman2000} & 8\,700 & 4.0 & \\
\citet{Sokolov1998} & 8\,850(450) & & \\
\hline
This work & 8\,900(200) & 3.9(2) & 1.582(15) \\ 
\hline
\hline
\end{tabular}
\end{center}
\end{table}

The used atmospheric models (plane-parallel, hydrostatic and radiative equilibrium, 1-dimensional) were calculated with the ATLAS\,9 code \citep{Kurucz2014}, whereas the synthetic spectra were computed with the line-blanketed, local thermodynamical equilibrium code SYNTHE \citep{Kurucz2005}. Both codes have been ported to GNU/Linux by \citet{Sbordone2005}. We used the latest version of line list available at the webpage of Fiorella Castelli\footnote{http://wwwuser.oats.inaf.it/castelli/}. 

The iron-peak elements, especially Cr, are strongly enhanced, whereas the light elements C and O are slightly underabundant. Sr is also strongly enhanced. We therefore conclude that 21\,Com is a typical CP2 object.

Finally, we searched for the presence of the Pr-Nd anomaly \citep{Ryabchikova2004}, which is a strong spectroscopic signature of the roAp phenomenon. The anomaly, reported for almost all roAp stars, manifests itself as a difference of at least 1.5 dex in element abundance derived separately from the lines of the second and first ions of Pr and Nd. The temperatures of the roAp stars analyzed in \citet{Ryabchikova2004} range from 6\,600\,K to 8\,100\,K. Our target, 21\,Com, is 800\,K hotter than the upper limit, which would render it an outstanding object among this group. We were only able to identify one unblended \ion{Nd}{II} line at 5248.79\,\AA. The abundance derived from this line does not differ from that derived from seven different, (partially-)blended \ion{Nd}{III} lines. Although based on only one unblended line, this demonstrates that 21 Com lacks one of the key characteristics of roAp stars, in line with the results from the frequency analysis (cf. Section \ref{section_periodanalysis}).

\subsection{Evolutionary models} \label{evol_models}

\begin{figure}
	\includegraphics[width=\columnwidth]{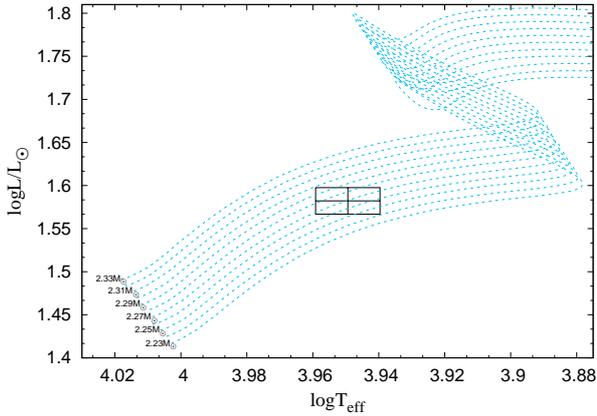}
	\caption{HR diagram with stellar evolutionary tracks, the position of 21\,Com, and its corresponding error box. Evolutionary models were calculated for masses from 2.23 to 2.33\,M$_{\odot}$ with a step-size of 0.02\,M$_{\odot}$. We assumed metallicity $Z=0.017$, initial hydrogen abundance $X_{\rm {ini}}=0.7$, and exponential overshooting parameter $f_{\rm {ov}}=0.01$.}
	\label{fig:HR}
\end{figure}

21\,Com has an accurately determined effective temperature and parallax, which strongly confine its global parameters. In Fig. \ref{fig:HR}, we show the Hertzsprung-Russell (HR) diagram with the location of 21\,Com and the corresponding error box, using the effective temperature determined from our spectroscopic analysis, $\log{T_{\rm{eff}}}$\,=\,$3.9494(98)$. The luminosity has been calculated with the Gaia parallax $\pi=12.009(140)$\,mas and is equal to $\log{L/\rm {L}}_{\odot}$\,=\,$1.582(15)$. The bolometric correction, BC\,=\,$-0.048\pm0.029$, was adopted from \citet{Flower96}.

For a given value of effective temperature and luminosity we can derive the stellar radius, $R$. Using rotational frequency and radius, the rotational velocity, $V_{\rm rot}$, can be obtained. Both parameters, $R$ and $V_{\rm{rot}}$, depend slightly on the position of the star in the HR diagram. In the observed error box, we found that the radius changes from 2.4 to 2.8\,R$_{\odot}$ and the rotational velocity from 60 to 68\,km\,s$^{-1}$. As a result, we have adopted $R$\,=\,$2.6\pm0.2$\,\rm{R}$_{\odot}$ and $V_{\rm {rot}}$\,=\,$64\pm4$\,km\,s$^{-1}$. With the determined value of $\vsini$\,=\,$63\pm2$ km\,s$^{-1}$ (cf. Section \ref{abundance_analysis}), we can also constrain the inclination angle $i\in\,\langle64$\textdegree$,90$\textdegree$\rangle$.

Evolutionary tracks for masses from 2.23 to 2.33\,M$_{\odot}$ were calculated with the Modules for Experiments in Stellar Astrophysics (MESA) code \citep{MESA,MESA2,MESA3,MESA4}. We also applied the MESA Isochrones and Stellar Tracks (MIST) configuration files \citep{MIST0,MIST1} and assumed metallicity $Z=0.017$, initial hydrogen abundance $X_{\rm{ini}}=0.7$ and exponential overshooting parameter from the hydrogen-burning convective core, $f_{\rm{ov}}=0.01$. In our computations, we used the OPAL opacity tables \citep{OPAL}, supplemented with the data provided by \cite{2005ApJ...623..585F} for the low temperature 
region. We employed the solar chemical element mixture as determined by \citet{AGSS09}.

From the HR diagram (Fig. \ref{fig:HR}), we can see that the mass of the star is of the order of 2.27\,M$_{\odot}$. Its position furthermore indicates that the star is most probably in the Main Sequence (MS) stage. Nevertheless, post-MS phases cannot be entirely ruled out. If we assume low metallicity of the order of 0.01, we can -- in the error box -- find models on the contraction phase (CoP) and after the loop (AL). Evolutionary changes of the rotational frequency are too small for detection, even for the fastest evolutionary phase ($df_{\rm {rot}}/dt\approx-1.1\times10^{-9}$ d$^{-1}$s$^{-1}$ on the MS, $2.8\times10^{-8}$ d$^{-1}$s$^{-1}$ on the contraction phase and $-4.6\times10^{-9}$ d$^{-1}$s$^{-1}$ after the loop). Therefore, we decided to check models on these different evolutionary phases. Theoretically, the star could be also in an early phase of evolution before the zero-age MS stage. However, 21\,Com is most certainly a member of the intermediate age open cluster Melotte 111, which excludes the possibility of its being a pre-MS star.  In the following, we will first consider the most probable case, viz. that the star is on the MS.

In order to constrain as many parameters as possible, we fitted the observed values of effective temperature $\log{T_{\mathrm{eff}}}=3.9494$ (8\,900 K), luminosity $\log{L/\rm {L}}_{\odot}$\,=\,$1.582$ and rotational frequency, $f_{\rm {rot}}$\,=\,$0.487$\,d$^{-1}$. Fitting $f_{\rm {rot}}$ required the simultaneous adjustment of rotational velocity and radius.

The MS models, called MS1 and MS2, were calculated for initial hydrogen abundance $X_{\rm {ini}}=0.700$, radius $R=2.6\,R_{\sun}$, rotational velocity $V_{\rm {rot}}=64$ km\,s$^{-1}$ and two values of metallicity, $Z=0.017$ and 0.020, respectively. Model MS3 has $Z=0.017$ and lower $X_{\rm {ini}}=0.650$, yielding a higher helium content. The parameters of all models are given in Table\,\ref{tab:modelsMS}, which is organized as follows:
\begin{itemize}
\item Column 1: model name.
\item Column 2: exponential overshooting parameter, $f_{\rm {ov}}$.
\item Column 3: initial abundances of hydrogen, $X_{\rm{ini}}$.
\item Column 4: initial abundances of helium, $Y_{\rm{ini}}$.
\item Column 5: metallicity, $Z$. 
\item Column 6: stellar mass, $M$ [M$_{\odot}$].
\item Column 7: age, calculated from cloud collapse.
\item Column 8: surface gravity.
\item Column 9: central hydrogen content, $X_c$.
\end{itemize}

The different metallicity input values produce small but significant changes in model mass, with higher values of $Z$ resulting in larger masses, e.g. $M$\,=\,2.2708\,\rm{M}$_{\odot}$ for $Z$\,=\,$0.017$ and $M$\,=\,2.3666\,\rm{M}$_{\odot}$ for $Z$\,=\,$0.020$. The core hydrogen abundance also increases with metallicity, while the age of a model decreases with increasing $Z$. 
\begin{table*}
	\begin{center}
	\caption{Models of 21\,Com fitting the observed values of effective temperature, luminosity, and rotational frequency.}
	\label{tab:modelsMS}
	\begin{tabular}{ccccccccc}
		\hline
		\hline
		Model & $f{\rm {ov}}$ & $X_{\rm {ini} }$ & $Y_{\rm {ini}}$  & $Z$ & $M$ & age &  $\log{g}$ & $X_{\rm{c}}$  \\
		& & & & & $[\rm{M}_{\odot}]$ & [Myr]  &  \\		
		\hline	

        MS1 & 0.01 & 0.700 & 0.283 & 0.017 & 2.2708 & 514.359  & 3.9628 & 0.310  \\
        MS2 & 0.01 & 0.700 & 0.280 & 0.020 & 2.3666 & 444.908  & 3.9808 & 0.316 \\ 
		MS3 & 0.01 & 0.650 & 0.333 & 0.017 & 2.1257 & 493.800  & 3.9341 & 0.251 \\		
		CoP & 0.00 & 0.700 & 0.290 & 0.010 & 2.0471 & 670.407  & 3.9178 & 0.005 \\
		AL  & 0.00 & 0.700 & 0.290 & 0.010 & 1.9849 & 732.774 &  3.9043 & 0.000 \\			
   		OPLIB MS1 & 0.01 & 0.700 & 0.283 & 0.017 & 2.2856 & 498.929 & 3.9656 & 0.310 \\
		\hline
		\hline
	\end{tabular}
\end{center}
\end{table*}

The contraction phase after the MS lasts about 3\% of the MS lifetime. The parameters of the corresponding model, which we have termed CoP, are given in Table\,\ref{tab:modelsMS}. It has a smaller mass than the MS models, i.e. $M$\,=\,$2.0471\,\rm{M}_{\odot}$. The more evolved model situated after the loop (termed AL) has an even smaller mass, $M$\,=\,$1.9849\,\rm{M}_{\odot}$. The duration of this evolutionary phase is about 2\% of the MS stage. It is important to point out that these two models, CoP and AL, have smaller metallicity and overshooting from the convective core. This adjustment was necessary to place the corresponding evolutionary paths inside the observed error box.

Last, we checked the impact of different opacity tables on the calculated models by using the new Los Alamos data \citep[][OPLIB]{OPLIB1,OPLIB2}. The resulting model, OPLIB MS1, is rather similar to that derived with the OPAL data (see Table\,\ref{tab:modelsMS}).

\begin{figure}
	\includegraphics[width=\columnwidth]{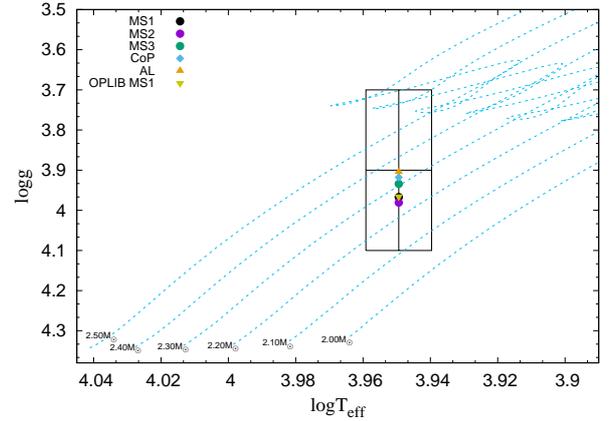}
	\caption{Kiel diagram with the position of 21\,Com as derived from spectroscopic analysis. The dotted lines are evolutionary tracks for
	the given masses. The abbreviations on the left hand side are for the different models of 21\,Com listed in Table\,\ref{tab:modelsMS}
	and discussed in Section \ref{evol_models}.}
	\label{fig:Kiel}
\end{figure}

A comparison of the theoretical gravity value with the one derived from spectroscopy gives rather good agreement. In Fig.\,\ref{fig:Kiel}, we show the Kiel diagram ($\log{T_{\rm{eff}}}$ vs. $\log{g}$) with the position of 21\,Com and the models from Table\,\ref{tab:modelsMS}. 

\subsection{Pulsational models} \label{pulsation_models}

The detection of short-term variability in the past has led to the tempting suggestion of the presence of pulsation in 21\,Com. However, judging from the available parameters, the star seems to be too hot for $\delta$ Scuti type pulsation, and the blue border of the theoretical instability strip in the HR diagram is situated to the right of the position of 21\,Com \citep[see for example][]{2000ASPC..210..215P,2004A&A...414L..17D,2016MNRAS.457.3163X}. Nevertheless, the instability region is not far from the parameters of 21\,Com and we decided to check this hypothesis by calculating pulsational models.

Theoretical frequencies were calculated with the customized non-adiabatic pulsational code of \cite{1977AcA....27...95D,Dziembowski1977}. The code uses the frozen convective flux approximation, which is not valid in the region with efficient convective energy transport. Fortunately, 21\,Com is hot enough to be appropriately calculated with this simplification because convective transport is not very efficient in our target star, as only about 5\% of energy will be transported by convective bubbles.

\begin{figure}
	\includegraphics[width=\columnwidth]{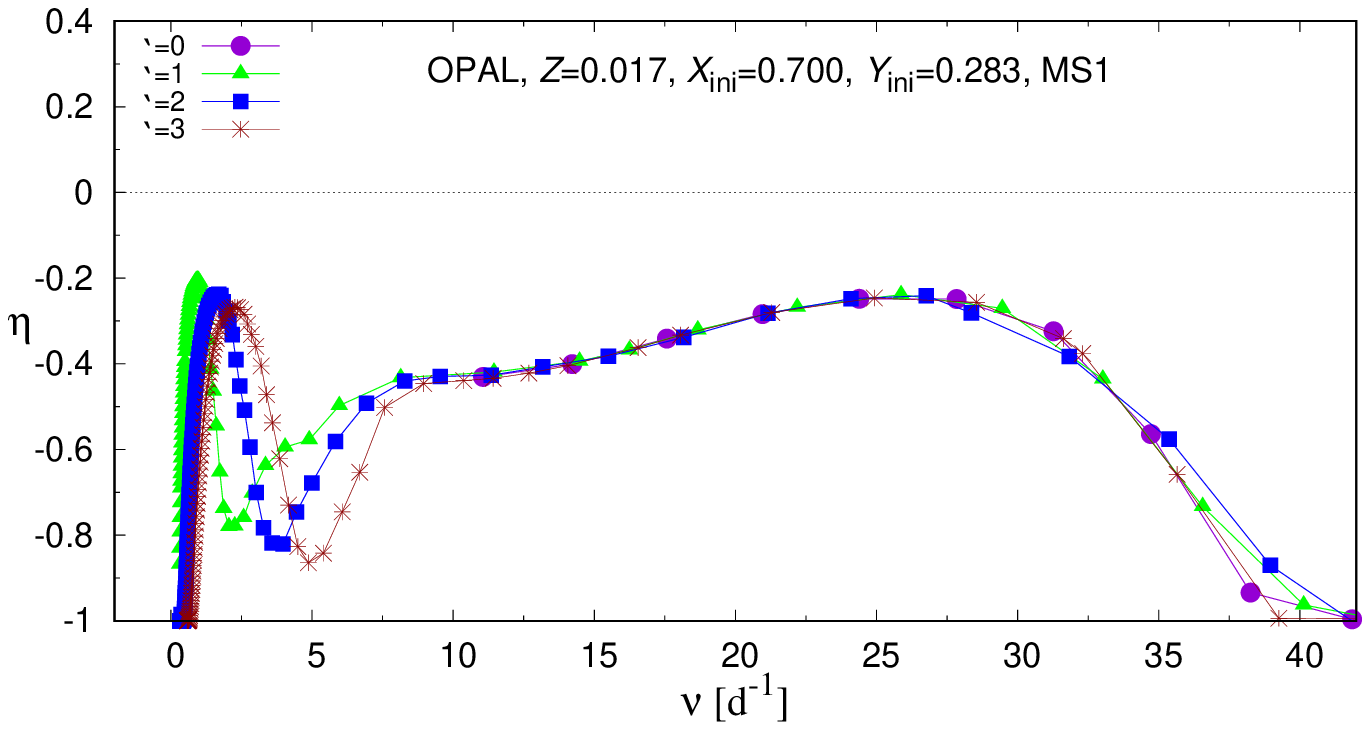}
	\includegraphics[width=\columnwidth]{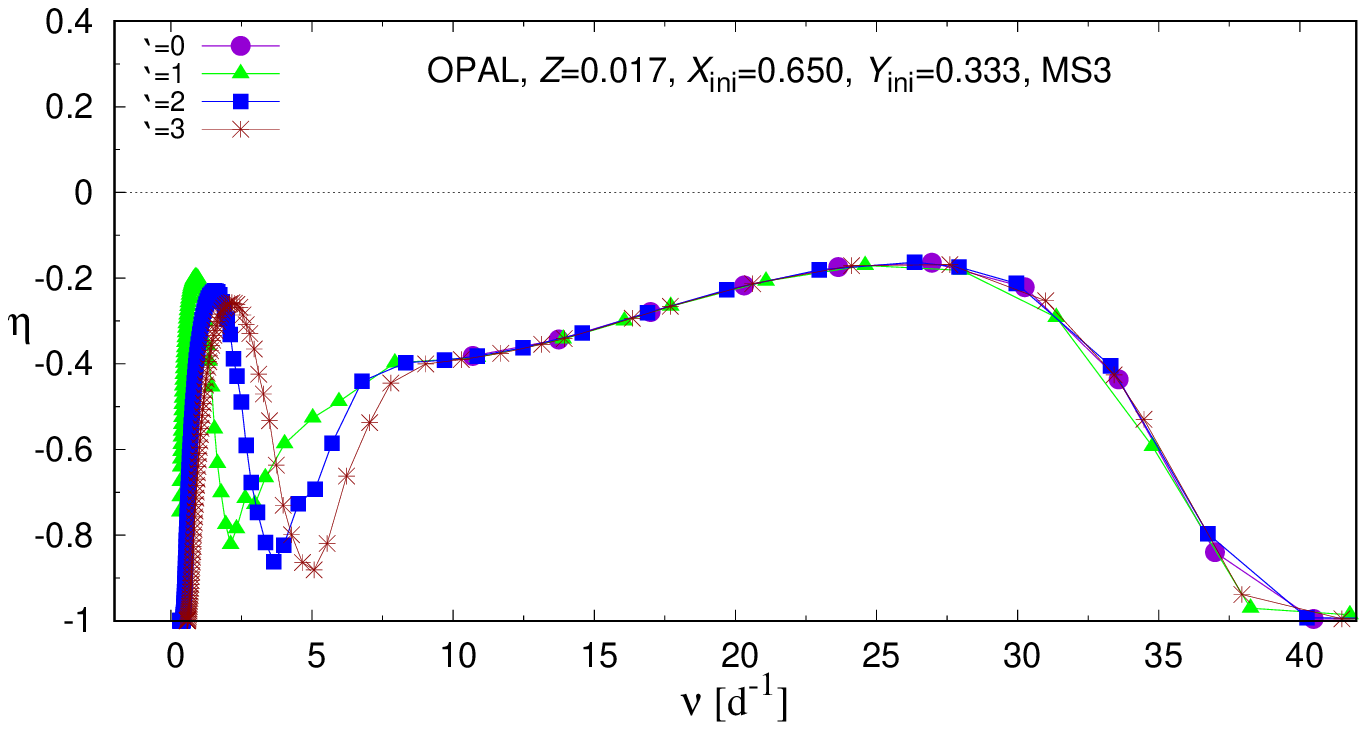}
	\caption{Instability parameter, $\eta$, as a function of frequency for the MS1 (upper panel) and MS3 (bottom panel) models of 21\,Com. Modes of degrees $\ell=0,1,2$ and 3 were considered.}
	\label{fig:nuetaMS}
\end{figure}

In Fig.\,\ref{fig:nuetaMS}, we plotted the instability parameter, $\eta$ \citep{Stellingwerf1978}, as a function of frequency for the MS1 (upper panel) and MS3 (bottom panel) models of 21\,Com. If $\eta>0$, then the corresponding mode is excited in a model. We have considered modes of degrees $\ell=0,1,2$ and 3. For clarity, only centroid modes, i.e. with azimuthal number $m=0$, are shown in the plots.

The models MS1 and MS3 differ primarily in assumed initial helium abundance, $Y=0.283$ for MS1 and $0.333$ for MS3. This is a very important parameter as the main driving mechanism for $\delta$ Scuti pulsation is the $\kappa$ mechanism operating in the He\,\textsc{II} partial ionization zone \citep{2005MNRAS.361..476D}. Due to this fact, we notice an increase in the instability parameter with increasing helium content. Nevertheless, modes are still stable throughout the observed frequency range. Including rotationally-split modes does not alter this situation because at $V_{\rm {rot}}$\,=\,$64\pm4$\,km\,s$^{-1}$, the resulting impact on the modes is negligible. The differences in metallicity between the models do not significantly impact the results either.

\begin{figure}
	\includegraphics[width=\columnwidth]{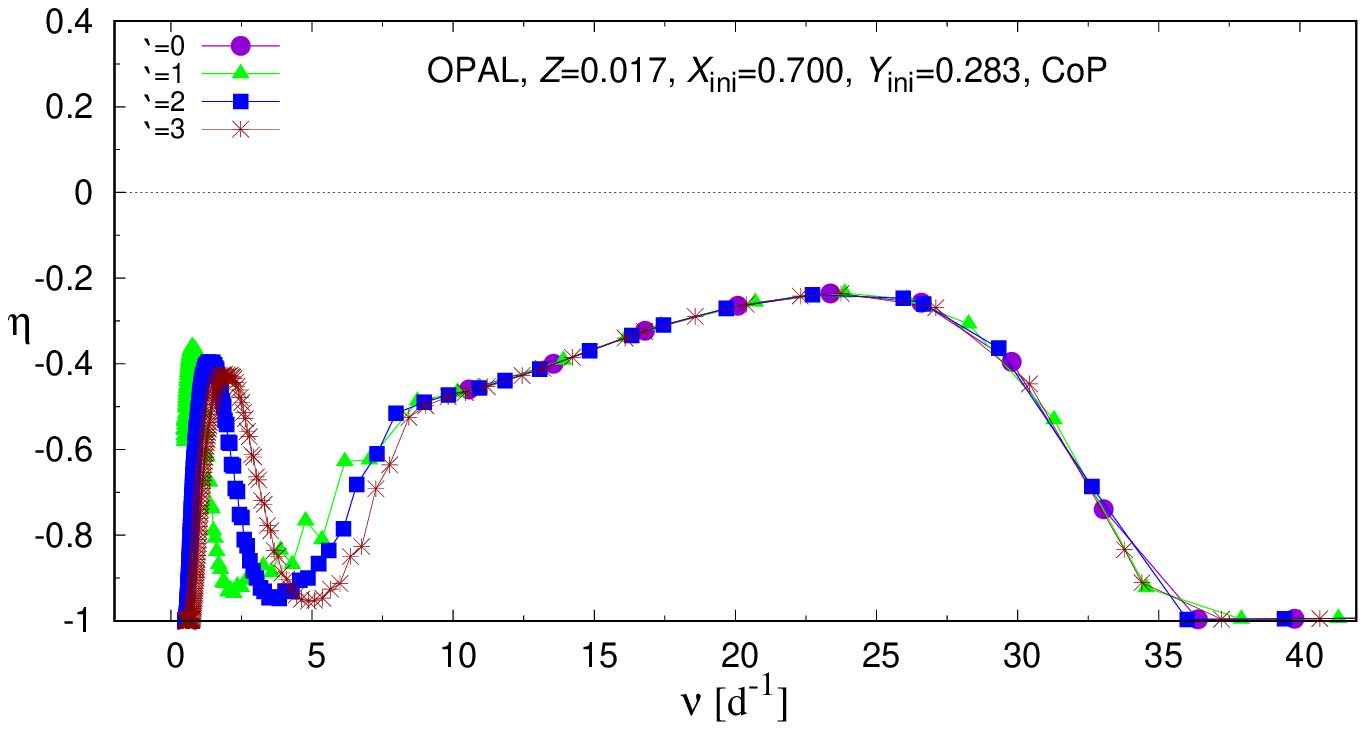}
	\includegraphics[width=\columnwidth]{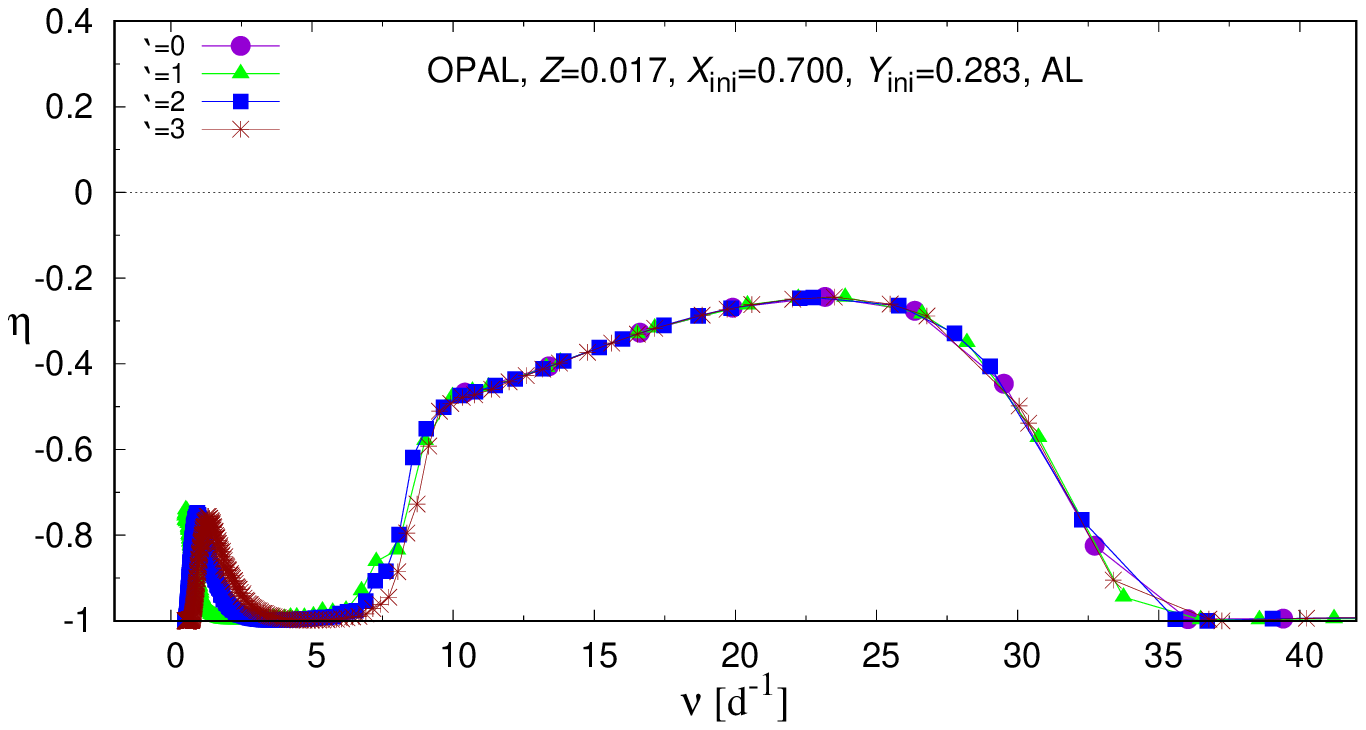}
	\caption{The same as in Fig.\,\ref{fig:nuetaMS}, but for the CoP (upper panel) and AL (bottom panel) models of 21\,Com.}
	\label{fig:nuetanonMS}
\end{figure}

The pulsational properties of the models of other evolutionary phases differ considerably. This can be clearly seen in Fig.\,\ref{fig:nuetanonMS}, which illustrates the instability parameter for the  CoP (middle panel) and AL (bottom panel) models. However, in no case, unstable modes were found.

In summary, our models do not predict the occurrence of unstable modes in the investigated frequency range, in agreement with the here presented null results from photometry and RV data. Without a significant increase of the opacity in the He\,\textsc{II} ionization zone, no unstable modes can be derived for the observed parameters of 21\,Com.

\section{Concluding remarks} \label{Conclusion}

Over four decades, the characteristics of the photometric and spectroscopic variability of 21\,Com have been a matter of debate. A consensus has only been reached on the large amplitude variability due to chemical abundance inhomogeneities (ACV variations), albeit different (though, in most cases, related; cf. Sect. \ref{history}) periods were published. The existence of $\delta$ Scuti and roAp pulsations, however, has been a debated topic throughout much of the star's observational history, which lead to the publication of many different periods and conflicting results. For the first time in 25 years, we have performed a new investigation of 21\,Com using state-of-the-art instrumentation (satellite and high-cadence ground-based photometry, time series spectroscopy) and pulsational modeling to shed more light on this matter.

Our abundance analysis confirms that 21\,Com is a classical CP2 star that shows increased abundances of, in particular, the iron-peak elements (most noteworthy Cr), and Sr. Our photometric data confirm the existence of rotational modulation on a period of 2.05219(2)\,d with an amplitude of 19\,mmag due to surface abundance inhomogeneities. However, no indication for additional variability on short time scales was found in our photometric observations and RV data. From the characteristics of our data, we estimate that, in the range from 5 to 399\,d$^{-1}$, frequencies with semi-amplitudes as low as 0.2\,mmag would have been detectable (cf. Section \ref{section_periodanalysis}). Thus, the data used here are perfectly suitable for the detection of the targeted $\delta$ Scuti and roAp pulsations. In summary, we conclude that, in the accuracy limit and time span of our extensive observational data, no short-term variability was present in 21\,Com.

18\,Com, which has been regularly employed as comparison star in the past, turned out to exhibit photometric variability with a period of $P$\,=\,1.41645\,d and an amplitude of 8\,mmag. However, this fact (alone) cannot explain the various literature claims as to the existence of photometric variability in 21\,Com on the time scales of minutes to two hours.

From spectroscopic analysis, we have derived $T_{\rm eff}\,=\,8\,900\pm200$\,K, $\log g\,=\,3.9\pm0.2$, and $\vsini\,=\,63\pm2$ km\,s$^{-1}$. Employing the Gaia parallax, we have calculated a stellar luminosity of $\log{L/\rm{L_{\sun} } }\,=\,1.582(15)$ and constrained the radius to $R$\,=\,$2.6\pm0.2$\,\rm{R}$_{\odot}$. Using the observed value of the rotational frequency, we furthermore derived a rotational velocity of $V_{\rm {rot}}$\,=\,$64\pm4$\,km\,s$^{-1}$. From $V_{\rm {rot}}$ and the observed value of \vsini, we conclude that the star is seen nearly equator-on, $i\in\,\langle64$\textdegree$,90$\textdegree$\rangle$. The stellar mass strongly depends on the employed model parameters. Assuming the most likely scenario that 21\,Com is a MS object, we have derived $M$\,=\,$2.29\pm0.10$\,M$_{\odot}$. None of our pulsational models predict the occurrence of unstable modes in the investigated frequency range, which is in agreement with the here presented null results from the time-series data.

While it is impossible to assess whether the star has exhibited short-term variability at other times/phases in the past, the new state-of-the-art and highly precise observational data, theoretical considerations (pulsational modeling) as well as the several issues/inconsistencies that have been identified in previous studies aimed at identifying short period variability strongly suggest that 21\,Com is neither a $\delta$ Scuti nor a roAp pulsator but rather a classical ``well-behaved'' CP2 star exhibiting rotational ACV variability.

\section*{Acknowledgments}

We thank the referee for the thoughtful report that helped to improve the paper. This work was financially supported by the Polish National Science Centre grants 2013/08/S/ST9/00583 and 2015/17/B/ST9/02082. Calculations have been partly carried out using resources provided by the Wroclaw Centre for Networking and Supercomputing (http://www.wcss.pl), grants No. 214 and 265. EN and TR are supported by NCN through the research grant No. 2014/13/B/ST9/00902. GH acknowledges support from the Polish National Science Center (NCN), grant no. 2015/18/A/ST9/00578. The research of M.F. was supported by the Slovak Research and Development Agency under the contract No. APVV-15-0458 and internal grant No. VVGS-PF-2018-758 of the Faculty of Science, P. J. \v{S}af\'{a}rik University in Ko\v{s}ice. This work has made use of data from the European Space Agency (ESA) mission {\it Gaia} (\url{https://www.cosmos.esa.int/gaia}), processed by the {\it Gaia} Data Processing and Analysis Consortium (DPAC, \url{https://www.cosmos.esa.int/web/gaia/dpac/consortium}). Funding for the DPAC has been provided by national institutions, in particular the institutions participating in the {\it Gaia} Multilateral Agreement.

\bibliographystyle{mnras}
\bibliography{21Com}

\begin{thebibliography}{}
\makeatletter
\relax
\def\mn@urlcharsother{\let\do\@makeother \do\$\do\&\do\#\do\^\do\_\do\%\do\~}
\def\mn@doi{\begingroup\mn@urlcharsother \@ifnextchar [ {\mn@doi@}
  {\mn@doi@[]}}
\def\mn@doi@[#1]#2{\def\@tempa{#1}\ifx\@tempa\@empty \href
  {http://dx.doi.org/#2} {doi:#2}\else \href {http://dx.doi.org/#2} {#1}\fi
  \endgroup}
\def\mn@eprint#1#2{\mn@eprint@#1:#2::\@nil}
\def\mn@eprint@arXiv#1{\href {http://arxiv.org/abs/#1} {{\tt arXiv:#1}}}
\def\mn@eprint@dblp#1{\href {http://dblp.uni-trier.de/rec/bibtex/#1.xml}
  {dblp:#1}}
\def\mn@eprint@#1:#2:#3:#4\@nil{\def\@tempa {#1}\def\@tempb {#2}\def\@tempc
  {#3}\ifx \@tempc \@empty \let \@tempc \@tempb \let \@tempb \@tempa \fi \ifx
  \@tempb \@empty \def\@tempb {arXiv}\fi \@ifundefined
  {mn@eprint@\@tempb}{\@tempb:\@tempc}{\expandafter \expandafter \csname
  mn@eprint@\@tempb\endcsname \expandafter{\@tempc}}}

\bibitem[\protect\citeauthoryear{{Abt} \& {Levy}}{{Abt} \&
  {Levy}}{1976}]{Abt1976}
{Abt} H.~A.,  {Levy} S.~G.,  1976, \mn@doi [\apjs] {10.1086/190363}, \href
  {http://adsabs.harvard.edu/abs/1976ApJS...30..273A} {30, 273}

\bibitem[\protect\citeauthoryear{{Abt} \& {Willmarth}}{{Abt} \&
  {Willmarth}}{1999}]{Abt1999}
{Abt} H.~A.,  {Willmarth} D.~W.,  1999, \mn@doi [\apj] {10.1086/307569}, \href
  {http://esoads.eso.org/abs/1999ApJ...521..682A} {521, 682}

\bibitem[\protect\citeauthoryear{{Adelman} \& {Rayle}}{{Adelman} \&
  {Rayle}}{2000}]{Adelman2000}
{Adelman} S.~J.,  {Rayle} K.~E.,  2000, \aap, \href
  {https://ui.adsabs.harvard.edu/\#abs/2000A&A...355..308A} {355, 308}

\bibitem[\protect\citeauthoryear{{Aslanov}, {Rustamov}, {Khalilov}  \&
  {Shakirzade}}{{Aslanov} et~al.}{1976}]{1976SvA....19..734A}
{Aslanov} I.~A.,  {Rustamov} I.~S.,  {Khalilov} V.~M.,   {Shakirzade} A.~A.,
  1976, \sovast, \href {http://esoads.eso.org/abs/1976SvA....19..734A} {19,
  734}

\bibitem[\protect\citeauthoryear{{Asplund}, {Grevesse}, {Sauval}  \&
  {Scott}}{{Asplund} et~al.}{2009}]{AGSS09}
{Asplund} M.,  {Grevesse} N.,  {Sauval} A.~J.,   {Scott} P.,  2009, \mn@doi
  [ARAA] {10.1146/annurev.astro.46.060407.145222}, \href
  {http://adsabs.harvard.edu/abs/2009ARA%26A..47..481A} {47, 481}

\bibitem[\protect\citeauthoryear{{Bagnulo}, {Landstreet}, {Fossati}  \&
  {Kochukhov}}{{Bagnulo} et~al.}{2012}]{Bagnulo2012}
{Bagnulo} S.,  {Landstreet} J.~D.,  {Fossati} L.,   {Kochukhov} O.,  2012,
  \mn@doi [\aap] {10.1051/0004-6361/201118098}, \href
  {http://esoads.eso.org/abs/2012A%26A...538A.129B} {538, A129}

\bibitem[\protect\citeauthoryear{{Bahner} \& {Mawridis}}{{Bahner} \&
  {Mawridis}}{1957}]{1957ZA.....41..254B}
{Bahner} K.,  {Mawridis} L.,  1957, \zap, \href
  {http://esoads.eso.org/abs/1957ZA.....41..254B} {41, 254}

\bibitem[\protect\citeauthoryear{{Balona}}{{Balona}}{2018}]{Balona18gaia}
{Balona} L.~A.,  2018, \mn@doi [\mnras] {10.1093/mnras/sty1511}, \href
  {http://adsabs.harvard.edu/abs/2018MNRAS.479..183B} {479, 183}

\bibitem[\protect\citeauthoryear{{Balona} et~al.,}{{Balona}
  et~al.}{2011a}]{2011MNRAS.410..517B}
{Balona} L.~A.,  et~al., 2011a, \mn@doi [\mnras]
  {10.1111/j.1365-2966.2010.17461.x}, \href
  {http://adsabs.harvard.edu/abs/2011MNRAS.410..517B} {410, 517}

\bibitem[\protect\citeauthoryear{{Balona}, {Guzik}, {Uytterhoeven}, {Smith},
  {Tenenbaum}  \& {Twicken}}{{Balona} et~al.}{2011b}]{Balona2011}
{Balona} L.~A.,  {Guzik} J.~A.,  {Uytterhoeven} K.,  {Smith} J.~C.,
  {Tenenbaum} P.,   {Twicken} J.~D.,  2011b, \mn@doi [\mnras]
  {10.1111/j.1365-2966.2011.18973.x}, \href
  {http://esoads.eso.org/abs/2011MNRAS.415.3531B} {415, 3531}

\bibitem[\protect\citeauthoryear{{Blanco} \& {Catalano}}{{Blanco} \&
  {Catalano}}{1972}]{1972AJ.....77..666B}
{Blanco} C.,  {Catalano} F.~A.,  1972, \mn@doi [\aj] {10.1086/111334}, \href
  {http://esoads.eso.org/abs/1972AJ.....77..666B} {77, 666}

\bibitem[\protect\citeauthoryear{{Blanco}, {Catalano}  \&
  {Strazzulla}}{{Blanco} et~al.}{1978}]{Blanco1978}
{Blanco} C.,  {Catalano} F.~A.,   {Strazzulla} G.,  1978, \aaps, \href
  {http://esoads.eso.org/abs/1978A%26AS...31..205B} {31, 205}

\bibitem[\protect\citeauthoryear{{Cannon} \& {Pickering}}{{Cannon} \&
  {Pickering}}{1920}]{Cannon1920}
{Cannon} A.~J.,  {Pickering} E.~C.,  1920, Annals of Harvard College
  Observatory, \href {http://adsabs.harvard.edu/abs/1920AnHar..95....1C} {95}

\bibitem[\protect\citeauthoryear{{Casagrande}, {Sch{\"o}nrich}, {Asplund},
  {Cassisi}, {Ram{\'{\i}}rez}, {Mel{\'e}ndez}, {Bensby}  \&
  {Feltzing}}{{Casagrande} et~al.}{2011}]{Casagrande11}
{Casagrande} L.,  {Sch{\"o}nrich} R.,  {Asplund} M.,  {Cassisi} S.,
  {Ram{\'{\i}}rez} I.,  {Mel{\'e}ndez} J.,  {Bensby} T.,   {Feltzing} S.,
  2011, \mn@doi [\aap] {10.1051/0004-6361/201016276}, \href
  {http://adsabs.harvard.edu/abs/2011A%26A...530A.138C} {530, A138}

\bibitem[\protect\citeauthoryear{{Catanzaro}, {Frasca}, {Molenda-{\.Z}akowicz}
  \& {Marilli}}{{Catanzaro} et~al.}{2010}]{Catanzaro2010}
{Catanzaro} G.,  {Frasca} A.,  {Molenda-{\.Z}akowicz} J.,   {Marilli} E.,
  2010, \mn@doi [\aap] {10.1051/0004-6361/201014189}, \href
  {http://esoads.eso.org/abs/2010A%26A...517A...3C} {517, A3}

\bibitem[\protect\citeauthoryear{{Choi}, {Dotter}, {Conroy}, {Cantiello},
  {Paxton}  \& {Johnson}}{{Choi} et~al.}{2016}]{MIST1}
{Choi} J.,  {Dotter} A.,  {Conroy} C.,  {Cantiello} M.,  {Paxton} B.,
  {Johnson} B.~D.,  2016, \mn@doi [ApJ] {10.3847/0004-637X/823/2/102}, \href
  {http://adsabs.harvard.edu/abs/2016ApJ...823..102C} {823, 102}

\bibitem[\protect\citeauthoryear{{Colgan}, {Kilcrease}, {Magee}, {Abdallah},
  {Sherrill}, {Fontes}, {Hakel}  \& {Zhang}}{{Colgan} et~al.}{2015}]{OPLIB2}
{Colgan} J.,  {Kilcrease} D.~P.,  {Magee} N.~H.,  {Abdallah} J.,  {Sherrill}
  M.~E.,  {Fontes} C.~J.,  {Hakel} P.,   {Zhang} H.~L.,  2015, \mn@doi [High
  Energy Density Physics] {10.1016/j.hedp.2015.02.006}, \href
  {http://adsabs.harvard.edu/abs/2015HEDP...14...33C} {14, 33}

\bibitem[\protect\citeauthoryear{{Colgan} et~al.,}{{Colgan}
  et~al.}{2016}]{OPLIB1}
{Colgan} J.,  et~al., 2016, \mn@doi [\apj] {10.3847/0004-637X/817/2/116}, \href
  {http://adsabs.harvard.edu/abs/2016ApJ...817..116C} {817, 116}

\bibitem[\protect\citeauthoryear{{Deutsch}}{{Deutsch}}{1955}]{1955PASP...67..342D}
{Deutsch} A.~J.,  1955, \mn@doi [\pasp] {10.1086/126839}, \href
  {http://esoads.eso.org/abs/1955PASP...67..342D} {67, 342}

\bibitem[\protect\citeauthoryear{{Deutsch}}{{Deutsch}}{1958}]{Deutsch58}
{Deutsch} A.~J.,  1958, in {Lehnert} B.,  ed.,  IAU Symposium Vol. 6,
  Electromagnetic Phenomena in Cosmical Physics. p.~209

\bibitem[\protect\citeauthoryear{{Donati}, {Catala}, {Landstreet}  \&
  {Petit}}{{Donati} et~al.}{2006}]{Donati2006}
{Donati} J.-F.,  {Catala} C.,  {Landstreet} J.~D.,   {Petit} P.,  2006, in
  {Casini} R.,  {Lites} B.~W.,  eds,  Astronomical Society of the Pacific
  Conference Series Vol. 358, Astronomical Society of the Pacific Conference
  Series. p.~362

\bibitem[\protect\citeauthoryear{{Dotter}}{{Dotter}}{2016}]{MIST0}
{Dotter} A.,  2016, \mn@doi [ApJS] {10.3847/0067-0049/222/1/8}, \href
  {http://adsabs.harvard.edu/abs/2016ApJS..222....8D} {222, 8}

\bibitem[\protect\citeauthoryear{{Dukes} \& {Adelman}}{{Dukes} \&
  {Adelman}}{2018}]{Dukes2018}
{Dukes} Jr. R.~J.,  {Adelman} S.~J.,  2018, \mn@doi [\pasp]
  {10.1088/1538-3873/aaa952}, \href
  {http://esoads.eso.org/abs/2018PASP..130d4202D} {130, 044202}

\bibitem[\protect\citeauthoryear{{Dupret}, {Grigahc{\`e}ne}, {Garrido},
  {Gabriel}  \& {Scuflaire}}{{Dupret} et~al.}{2004}]{2004A&A...414L..17D}
{Dupret} M.-A.,  {Grigahc{\`e}ne} A.,  {Garrido} R.,  {Gabriel} M.,
  {Scuflaire} R.,  2004, \mn@doi [\aap] {10.1051/0004-6361:20031740}, \href
  {http://adsabs.harvard.edu/abs/2004A%26A...414L..17D} {414, L17}

\bibitem[\protect\citeauthoryear{{Dupret}, {Grigahc{\`e}ne}, {Garrido}, {De
  Ridder}, {Scuflaire}  \& {Gabriel}}{{Dupret}
  et~al.}{2005}]{2005MNRAS.361..476D}
{Dupret} M.-A.,  {Grigahc{\`e}ne} A.,  {Garrido} R.,  {De Ridder} J.,
  {Scuflaire} R.,   {Gabriel} M.,  2005, \mn@doi [\mnras]
  {10.1111/j.1365-2966.2005.09187.x}, \href
  {http://adsabs.harvard.edu/abs/2005MNRAS.361..476D} {361, 476}

\bibitem[\protect\citeauthoryear{{Dziembowski}}{{Dziembowski}}{1977a}]{1977AcA....27...95D}
{Dziembowski} W.,  1977a, \actaa, \href
  {http://adsabs.harvard.edu/abs/1977AcA....27...95D} {27, 95}

\bibitem[\protect\citeauthoryear{{Dziembowski}}{{Dziembowski}}{1977b}]{Dziembowski1977}
{Dziembowski} W.,  1977b, \actaa, \href
  {http://esoads.eso.org/abs/1977AcA....27..203D} {27, 203}

\bibitem[\protect\citeauthoryear{{Escorza} et~al.,}{{Escorza}
  et~al.}{2016}]{Escorza2016}
{Escorza} A.,  et~al., 2016, \mn@doi [\aap] {10.1051/0004-6361/201527870},
  \href {http://esoads.eso.org/abs/2016A%26A...588A..71E} {588, A71}

\bibitem[\protect\citeauthoryear{{Ferguson}, {Alexander}, {Allard}, {Barman},
  {Bodnarik}, {Hauschildt}, {Heffner-Wong}  \& {Tamanai}}{{Ferguson}
  et~al.}{2005}]{2005ApJ...623..585F}
{Ferguson} J.~W.,  {Alexander} D.~R.,  {Allard} F.,  {Barman} T.,  {Bodnarik}
  J.~G.,  {Hauschildt} P.~H.,  {Heffner-Wong} A.,   {Tamanai} A.,  2005,
  \mn@doi [ApJ] {10.1086/428642}, \href
  {http://adsabs.harvard.edu/abs/2005ApJ...623..585F} {623, 585}

\bibitem[\protect\citeauthoryear{{Flower}}{{Flower}}{1996}]{Flower96}
{Flower} P.~J.,  1996, \mn@doi [\apj] {10.1086/177785}, \href
  {http://esoads.eso.org/abs/1996ApJ...469..355F} {469, 355}

\bibitem[\protect\citeauthoryear{{Gaia Collaboration} et~al.,}{{Gaia
  Collaboration} et~al.}{2016}]{Gaia16}
{Gaia Collaboration} et~al., 2016, \mn@doi [\aap]
  {10.1051/0004-6361/201629272}, \href
  {http://adsabs.harvard.edu/abs/2016A%26A...595A...1G} {595, A1}

\bibitem[\protect\citeauthoryear{{Gaia Collaboration} et~al.,}{{Gaia
  Collaboration} et~al.}{2018}]{Gaia18}
{Gaia Collaboration} et~al., 2018, \mn@doi [\aap]
  {10.1051/0004-6361/201833051}, \href
  {http://adsabs.harvard.edu/abs/2018A%26A...616A...1G} {616, A1}

\bibitem[\protect\citeauthoryear{{Garrido} \& {Sanchez-Lavega}}{{Garrido} \&
  {Sanchez-Lavega}}{1983}]{1983IBVS.2368....1G}
{Garrido} R.,  {Sanchez-Lavega} A.,  1983, Information Bulletin on Variable
  Stars, \href {http://esoads.eso.org/abs/1983IBVS.2368....1G} {2368}

\bibitem[\protect\citeauthoryear{{Gebran}, {Farah}, {Paletou}, {Monier}  \&
  {Watson}}{{Gebran} et~al.}{2016}]{Gebran2016}
{Gebran} M.,  {Farah} W.,  {Paletou} F.,  {Monier} R.,   {Watson} V.,  2016,
  \mn@doi [\aap] {10.1051/0004-6361/201528052}, \href
  {https://ui.adsabs.harvard.edu/\#abs/2016A&A...589A..83G} {589, A83}

\bibitem[\protect\citeauthoryear{{Goncharskii}, {Ryabchikova}, {Stepanov},
  {Khokhlova}  \& {Yagola}}{{Goncharskii} et~al.}{1983}]{Goncharskii83}
{Goncharskii} A.~V.,  {Ryabchikova} T.~A.,  {Stepanov} V.~V.,  {Khokhlova}
  V.~L.,   {Yagola} A.~G.,  1983, \sovast, \href
  {http://adsabs.harvard.edu/abs/1983SvA....27...49G} {27, 49}

\bibitem[\protect\citeauthoryear{{Handler} \& {Shobbrook}}{{Handler} \&
  {Shobbrook}}{2002}]{Handler02}
{Handler} G.,  {Shobbrook} R.~R.,  2002, \mn@doi [\mnras]
  {10.1046/j.1365-8711.2002.05401.x}, \href
  {http://adsabs.harvard.edu/abs/2002MNRAS.333..251H} {333, 251}

\bibitem[\protect\citeauthoryear{{H{\"u}mmerich}, {Bernhard}, {Paunzen},
  {Hambsch}, {Bohlsen}  \& {Powles}}{{H{\"u}mmerich}
  et~al.}{2017}]{2017MNRAS.466.1399H}
{H{\"u}mmerich} S.,  {Bernhard} K.,  {Paunzen} E.,  {Hambsch} F.-J.,  {Bohlsen}
  T.,   {Powles} J.,  2017, \mn@doi [\mnras] {10.1093/mnras/stw3186}, \href
  {http://adsabs.harvard.edu/abs/2017MNRAS.466.1399H} {466, 1399}

\bibitem[\protect\citeauthoryear{{Iglesias} \& {Rogers}}{{Iglesias} \&
  {Rogers}}{1996}]{OPAL}
{Iglesias} C.~A.,  {Rogers} F.~J.,  1996, \mn@doi [ApJ] {10.1086/177381}, \href
  {http://adsabs.harvard.edu/abs/1996ApJ...464..943I} {464, 943}

\bibitem[\protect\citeauthoryear{{Jarzebowski}}{{Jarzebowski}}{1982}]{1982CoKon..83..190J}
{Jarzebowski} T.,  1982, Commmunications of the Konkoly Observatory Hungary,
  \href {http://esoads.eso.org/abs/1982CoKon..83..190J} {83, 190}

\bibitem[\protect\citeauthoryear{{Kharchenko}, {Piskunov}, {R{\"o}ser},
  {Schilbach}  \& {Scholz}}{{Kharchenko} et~al.}{2004}]{Kharchenko04}
{Kharchenko} N.~V.,  {Piskunov} A.~E.,  {R{\"o}ser} S.,  {Schilbach} E.,
  {Scholz} R.-D.,  2004, \mn@doi [Astronomische Nachrichten]
  {10.1002/asna.200410256}, \href
  {http://adsabs.harvard.edu/abs/2004AN....325..740K} {325, 740}

\bibitem[\protect\citeauthoryear{{Kochukhov} \& {Bagnulo}}{{Kochukhov} \&
  {Bagnulo}}{2006}]{Kochukhov2006}
{Kochukhov} O.,  {Bagnulo} S.,  2006, \mn@doi [\aap]
  {10.1051/0004-6361:20054596}, \href
  {https://ui.adsabs.harvard.edu/\#abs/2006A&A...450..763K} {450, 763}

\bibitem[\protect\citeauthoryear{{Kochukhov}, {Shultz}  \&
  {Neiner}}{{Kochukhov} et~al.}{2019}]{Kochukhov19}
{Kochukhov} O.,  {Shultz} M.,   {Neiner} C.,  2019, \mn@doi [\aap]
  {10.1051/0004-6361/201834279}, \href
  {http://adsabs.harvard.edu/abs/2019A%26A...621A..47K} {621, A47}

\bibitem[\protect\citeauthoryear{{Koleva} \& {Vazdekis}}{{Koleva} \&
  {Vazdekis}}{2012}]{Koleva2012}
{Koleva} M.,  {Vazdekis} A.,  2012, \mn@doi [\aap]
  {10.1051/0004-6361/201118065}, \href
  {https://ui.adsabs.harvard.edu/\#abs/2012A&A...538A.143K} {538, A143}

\bibitem[\protect\citeauthoryear{{Kreidl} et~al.,}{{Kreidl}
  et~al.}{1990}]{1990MNRAS.245..642K}
{Kreidl} T.~J.,  et~al., 1990, \mnras, \href
  {http://esoads.eso.org/abs/1990MNRAS.245..642K} {245, 642}

\bibitem[\protect\citeauthoryear{{Krti{\v{c}}ka}, {Jan{\'\i}k}, {Markov{\'a}},
  {Mikul{\'a}{\v{s}}ek}, {Zverko}, {Prv{\'a}k}  \& {Skarka}}{{Krti{\v{c}}ka}
  et~al.}{2013}]{Krticka2013}
{Krti{\v{c}}ka} J.,  {Jan{\'\i}k} J.,  {Markov{\'a}} H.,  {Mikul{\'a}{\v{s}}ek}
  Z.,  {Zverko} J.,  {Prv{\'a}k} M.,   {Skarka} M.,  2013, \mn@doi [\aap]
  {10.1051/0004-6361/201221018}, \href
  {https://ui.adsabs.harvard.edu/#abs/2013A&A...556A..18K} {556, A18}

\bibitem[\protect\citeauthoryear{{Kurtz}}{{Kurtz}}{1982}]{1982MNRAS.200..807K}
{Kurtz} D.~W.,  1982, \mn@doi [\mnras] {10.1093/mnras/200.3.807}, \href
  {http://adsabs.harvard.edu/abs/1982MNRAS.200..807K} {200, 807}

\bibitem[\protect\citeauthoryear{{Kurtz}, {Hubrig}, {Gonz{\'a}lez}, {van Wyk}
  \& {Martinez}}{{Kurtz} et~al.}{2008}]{Kurtz2008}
{Kurtz} D.~W.,  {Hubrig} S.,  {Gonz{\'a}lez} J.~F.,  {van Wyk} F.,   {Martinez}
  P.,  2008, \mn@doi [\mnras] {10.1111/j.1365-2966.2008.13154.x}, \href
  {http://esoads.eso.org/abs/2008MNRAS.386.1750K} {386, 1750}

\bibitem[\protect\citeauthoryear{{Kurucz}}{{Kurucz}}{2005}]{Kurucz2005}
{Kurucz} R.~L.,  2005, Memorie della Societa Astronomica Italiana Supplementi,
  \href {http://esoads.eso.org/abs/2005MSAIS...8...14K} {8, 14}

\bibitem[\protect\citeauthoryear{{Kurucz}}{{Kurucz}}{2014}]{Kurucz2014}
{Kurucz} R.~L.,  2014, {Model Atmosphere Codes: ATLAS12 and ATLAS9}.
pp 39--51, \mn@doi{10.1007/978-3-319-06956-2_4}

\bibitem[\protect\citeauthoryear{{Lampens} et~al.,}{{Lampens}
  et~al.}{2013}]{Lampens2013}
{Lampens} P.,  et~al., 2013, \mn@doi [\aap] {10.1051/0004-6361/201219525},
  \href {http://esoads.eso.org/abs/2013A%26A...549A.104L} {549, A104}

\bibitem[\protect\citeauthoryear{{Landstreet} et~al.,}{{Landstreet}
  et~al.}{2008}]{Landstreet2008a}
{Landstreet} J.~D.,  et~al., 2008, \mn@doi [\aap] {10.1051/0004-6361:20078884},
  \href {https://ui.adsabs.harvard.edu/#abs/2008A&A...481..465L} {481, 465}

\bibitem[\protect\citeauthoryear{{Lenz} \& {Breger}}{{Lenz} \&
  {Breger}}{2005}]{period04}
{Lenz} P.,  {Breger} M.,  2005, \mn@doi [Communications in Asteroseismology]
  {10.1553/cia146s53}, \href
  {http://adsabs.harvard.edu/abs/2005CoAst.146...53L} {146, 53}

\bibitem[\protect\citeauthoryear{{Leone} \& {Catanzaro}}{{Leone} \&
  {Catanzaro}}{2001}]{Leone2001}
{Leone} F.,  {Catanzaro} G.,  2001, \mn@doi [\aap]
  {10.1051/0004-6361:20000450}, \href
  {http://esoads.eso.org/abs/2001A%26A...365..118L} {365, 118}

\bibitem[\protect\citeauthoryear{{Lipski} \& {St{\c e}pie{\'n}}}{{Lipski} \&
  {St{\c e}pie{\'n}}}{2008}]{Lipski2008}
{Lipski} {\L}.,  {St{\c e}pie{\'n}} K.,  2008, \mn@doi [\mnras]
  {10.1111/j.1365-2966.2008.12856.x}, \href
  {https://ui.adsabs.harvard.edu/\#abs/2008MNRAS.385..481L} {385, 481}

\bibitem[\protect\citeauthoryear{{Luri} et~al.,}{{Luri} et~al.}{2018}]{Luri18}
{Luri} X.,  et~al., 2018, \mn@doi [\aap] {10.1051/0004-6361/201832964}, \href
  {http://adsabs.harvard.edu/abs/2018A%26A...616A...9L} {616, A9}

\bibitem[\protect\citeauthoryear{{Maitzen}, {Albrecht}  \& {Heck}}{{Maitzen}
  et~al.}{1978}]{Maitzen1978}
{Maitzen} H.~M.,  {Albrecht} R.,   {Heck} A.,  1978, \aap, \href
  {http://esoads.eso.org/abs/1978A%26A....62..199M} {62, 199}

\bibitem[\protect\citeauthoryear{{Mantegazza} \& {Poretti}}{{Mantegazza} \&
  {Poretti}}{1999}]{Mantegazza1999}
{Mantegazza} L.,  {Poretti} E.,  1999, \aap, \href
  {http://adsabs.harvard.edu/abs/1999A%26A...348..139M} {348, 139}

\bibitem[\protect\citeauthoryear{{Mantegazza} \& {Poretti}}{{Mantegazza} \&
  {Poretti}}{2002}]{Mantegazza2002}
{Mantegazza} L.,  {Poretti} E.,  2002, \mn@doi [\aap]
  {10.1051/0004-6361:20021456}, \href
  {http://esoads.eso.org/abs/2002A%26A...396..911M} {396, 911}

\bibitem[\protect\citeauthoryear{{McDonald}, {Zijlstra}  \& {Boyer}}{{McDonald}
  et~al.}{2012}]{McDonald2012}
{McDonald} I.,  {Zijlstra} A.~A.,   {Boyer} M.~L.,  2012, \mn@doi [\mnras]
  {10.1111/j.1365-2966.2012.21873.x}, \href
  {https://ui.adsabs.harvard.edu/\#abs/2012MNRAS.427..343M} {427, 343}

\bibitem[\protect\citeauthoryear{{Michaud}, {Megessier}  \&
  {Charland}}{{Michaud} et~al.}{1981}]{Michaud1981}
{Michaud} G.,  {Megessier} C.,   {Charland} Y.,  1981, \aap, \href
  {https://ui.adsabs.harvard.edu/#abs/1981A&A...103..244M} {103, 244}

\bibitem[\protect\citeauthoryear{{Morris}}{{Morris}}{1985}]{Morris1985}
{Morris} S.~L.,  1985, \mn@doi [\apj] {10.1086/163359}, \href
  {http://adsabs.harvard.edu/abs/1985ApJ...295..143M} {295, 143}

\bibitem[\protect\citeauthoryear{{Musielok} \& {Kozar}}{{Musielok} \&
  {Kozar}}{1982}]{1982IBVS.2237....1M}
{Musielok} B.,  {Kozar} T.,  1982, Information Bulletin on Variable Stars,
  \href {http://esoads.eso.org/abs/1982IBVS.2237....1M} {2237}

\bibitem[\protect\citeauthoryear{{Neiner} \& {Lampens}}{{Neiner} \&
  {Lampens}}{2015}]{Neiner2015}
{Neiner} C.,  {Lampens} P.,  2015, \mn@doi [\mnras] {10.1093/mnrasl/slv130},
  \href {http://esoads.eso.org/abs/2015MNRAS.454L..86N} {454, L86}

\bibitem[\protect\citeauthoryear{{Neiner}, {Wade}  \& {Sikora}}{{Neiner}
  et~al.}{2017}]{Neiner2017}
{Neiner} C.,  {Wade} G.~A.,   {Sikora} J.,  2017, \mn@doi [\mnras]
  {10.1093/mnrasl/slx023}, \href
  {http://esoads.eso.org/abs/2017MNRAS.468L..46N} {468, L46}

\bibitem[\protect\citeauthoryear{{Nelson} \& {Kreidl}}{{Nelson} \&
  {Kreidl}}{1993}]{1993AJ....105.1903N}
{Nelson} M.~J.,  {Kreidl} T.~J.,  1993, \mn@doi [\aj] {10.1086/116565}, \href
  {http://esoads.eso.org/abs/1993AJ....105.1903N} {105, 1903}

\bibitem[\protect\citeauthoryear{{Netopil}, {Paunzen}, {Maitzen}, {North}  \&
  {Hubrig}}{{Netopil} et~al.}{2008}]{Netopil2008}
{Netopil} M.,  {Paunzen} E.,  {Maitzen} H.~M.,  {North} P.,   {Hubrig} S.,
  2008, \mn@doi [\aap] {10.1051/0004-6361:200810325}, \href
  {https://ui.adsabs.harvard.edu/\#abs/2008A&A...491..545N} {491, 545}

\bibitem[\protect\citeauthoryear{{Netopil}, {Paunzen}  \& {Carraro}}{{Netopil}
  et~al.}{2015}]{Netopil2015}
{Netopil} M.,  {Paunzen} E.,   {Carraro} G.,  2015, \mn@doi [\aap]
  {10.1051/0004-6361/201526372}, \href
  {http://adsabs.harvard.edu/abs/2015A%26A...582A..19N} {582, A19}

\bibitem[\protect\citeauthoryear{{Niemczura}, {Scholz}, {Hubrig},
  {J{\"a}rvinen}, {Sch{\"o}ller}, {Ilyin}  \& {Kahraman Ali{\c c}avu{\c
  s}}}{{Niemczura} et~al.}{2017}]{Niemczura2017}
{Niemczura} E.,  {Scholz} R.-D.,  {Hubrig} S.,  {J{\"a}rvinen} S.~P.,
  {Sch{\"o}ller} M.,  {Ilyin} I.,   {Kahraman Ali{\c c}avu{\c s}} F.,  2017,
  \mn@doi [\mnras] {10.1093/mnras/stx1377}, \href
  {http://esoads.eso.org/abs/2017MNRAS.470.3806N} {470, 3806}

\bibitem[\protect\citeauthoryear{{Oksala} et~al.,}{{Oksala}
  et~al.}{2015}]{Oksala2015}
{Oksala} M.~E.,  et~al., 2015, \mn@doi [\mnras] {10.1093/mnras/stv1086}, \href
  {http://adsabs.harvard.edu/abs/2015MNRAS.451.2015O} {451, 2015}

\bibitem[\protect\citeauthoryear{{Pamyatnykh}}{{Pamyatnykh}}{2000}]{2000ASPC..210..215P}
{Pamyatnykh} A.~A.,  2000, in {Breger} M.,  {Montgomery} M.,  eds,
  Astronomical Society of the Pacific Conference Series Vol. 210, Delta Scuti
  and Related Stars. p.~215 (\mn@eprint {} {astro-ph/0005276})

\bibitem[\protect\citeauthoryear{{Paunzen}, {Wraight}, {Fossati}, {Netopil},
  {White}  \& {Bewsher}}{{Paunzen} et~al.}{2013}]{2013MNRAS.429..119P}
{Paunzen} E.,  {Wraight} K.~T.,  {Fossati} L.,  {Netopil} M.,  {White} G.~J.,
  {Bewsher} D.,  2013, \mn@doi [\mnras] {10.1093/mnras/sts318}, \href
  {http://adsabs.harvard.edu/abs/2013MNRAS.429..119P} {429, 119}

\bibitem[\protect\citeauthoryear{{Paxton}, {Bildsten}, {Dotter}, {Herwig},
  {Lesaffre}  \& {Timmes}}{{Paxton} et~al.}{2011}]{MESA}
{Paxton} B.,  {Bildsten} L.,  {Dotter} A.,  {Herwig} F.,  {Lesaffre} P.,
  {Timmes} F.,  2011, \mn@doi [\apjs] {10.1088/0067-0049/192/1/3}, \href
  {http://esoads.eso.org/abs/2011ApJS..192....3P} {192, 3}

\bibitem[\protect\citeauthoryear{{Paxton} et~al.,}{{Paxton}
  et~al.}{2013}]{MESA2}
{Paxton} B.,  et~al., 2013, \mn@doi [\apjs] {10.1088/0067-0049/208/1/4}, \href
  {http://esoads.eso.org/abs/2013ApJS..208....4P} {208, 4}

\bibitem[\protect\citeauthoryear{{Paxton} et~al.,}{{Paxton}
  et~al.}{2015}]{MESA3}
{Paxton} B.,  et~al., 2015, \mn@doi [ApJS] {10.1088/0067-0049/220/1/15}, \href
  {http://adsabs.harvard.edu/abs/2015ApJS..220...15P} {220, 15}

\bibitem[\protect\citeauthoryear{{Paxton} et~al.,}{{Paxton}
  et~al.}{2018}]{MESA4}
{Paxton} B.,  et~al., 2018, \mn@doi [\apjs] {10.3847/1538-4365/aaa5a8}, \href
  {http://adsabs.harvard.edu/abs/2018ApJS..234...34P} {234, 34}

\bibitem[\protect\citeauthoryear{{Percy}}{{Percy}}{1973}]{1973AA....22..381P}
{Percy} J.~R.,  1973, \aap, \href
  {http://esoads.eso.org/abs/1973A%26A....22..381P} {22, 381}

\bibitem[\protect\citeauthoryear{{Percy}}{{Percy}}{1975}]{1975AJ.....80..698P}
{Percy} J.~R.,  1975, \mn@doi [\aj] {10.1086/111800}, \href
  {http://esoads.eso.org/abs/1975AJ.....80..698P} {80, 698}

\bibitem[\protect\citeauthoryear{{Pribulla} et~al.,}{{Pribulla}
  et~al.}{2015}]{Pribulla2015}
{Pribulla} T.,  et~al., 2015, \mn@doi [Astronomische Nachrichten]
  {10.1002/asna.201512202}, \href
  {http://adsabs.harvard.edu/abs/2015AN....336..682P} {336, 682}

\bibitem[\protect\citeauthoryear{{Provin}}{{Provin}}{1953}]{1953ApJ...118..489P}
{Provin} S.~S.,  1953, \mn@doi [\apj] {10.1086/145777}, \href
  {http://esoads.eso.org/abs/1953ApJ...118..489P} {118, 489}

\bibitem[\protect\citeauthoryear{{Pyper}}{{Pyper}}{1969}]{Pyper69}
{Pyper} D.~M.,  1969, \mn@doi [\apjs] {10.1086/190193}, \href
  {http://adsabs.harvard.edu/abs/1969ApJS...18..347P} {18, 347}

\bibitem[\protect\citeauthoryear{{Reegen} et~al.,}{{Reegen}
  et~al.}{2006}]{Reegen2006}
{Reegen} P.,  et~al., 2006, \mn@doi [\mnras]
  {10.1111/j.1365-2966.2006.10082.x}, \href
  {https://ui.adsabs.harvard.edu/#abs/2006MNRAS.367.1417R} {367, 1417}

\bibitem[\protect\citeauthoryear{{Reiners}}{{Reiners}}{2006}]{Reiners06}
{Reiners} A.,  2006, \mn@doi [\aap] {10.1051/0004-6361:20053911}, \href
  {http://adsabs.harvard.edu/abs/2006A%26A...446..267R} {446, 267}

\bibitem[\protect\citeauthoryear{{Renson} \& {Manfroid}}{{Renson} \&
  {Manfroid}}{2009}]{RM09}
{Renson} P.,  {Manfroid} J.,  2009, \mn@doi [\aap]
  {10.1051/0004-6361/200810788}, \href
  {http://adsabs.harvard.edu/abs/2009A%26A...498..961R} {498, 961}

\bibitem[\protect\citeauthoryear{{Robinson}, {Ammons}, {Kretke}, {Strader},
  {Wertheimer}, {Fischer}  \& {Laughlin}}{{Robinson} et~al.}{2007}]{Robinson07}
{Robinson} S.~E.,  {Ammons} S.~M.,  {Kretke} K.~A.,  {Strader} J.,
  {Wertheimer} J.~G.,  {Fischer} D.~A.,   {Laughlin} G.,  2007, \mn@doi [\apjs]
  {10.1086/513108}, \href {http://adsabs.harvard.edu/abs/2007ApJS..169..430R}
  {169, 430}

\bibitem[\protect\citeauthoryear{{Ryabchikova}, {Nesvacil}, {Weiss},
  {Kochukhov}  \& {St{\"u}tz}}{{Ryabchikova} et~al.}{2004}]{Ryabchikova2004}
{Ryabchikova} T.,  {Nesvacil} N.,  {Weiss} W.~W.,  {Kochukhov} O.,
  {St{\"u}tz} C.,  2004, \mn@doi [\aap] {10.1051/0004-6361:20041012}, \href
  {http://esoads.eso.org/abs/2004A%26A...423..705R} {423, 705}

\bibitem[\protect\citeauthoryear{{Santagati}, {Rodono}, {Ventura}  \&
  {Poretti}}{{Santagati} et~al.}{1989}]{1989IBVS.3373....1S}
{Santagati} G.,  {Rodono} M.,  {Ventura} R.,   {Poretti} E.,  1989, Information
  Bulletin on Variable Stars, \href
  {http://esoads.eso.org/abs/1989IBVS.3373....1S} {3373}

\bibitem[\protect\citeauthoryear{{Sbordone}}{{Sbordone}}{2005}]{Sbordone2005}
{Sbordone} L.,  2005, Memorie della Societa Astronomica Italiana Supplementi,
  \href {http://esoads.eso.org/abs/2005MSAIS...8...61S} {8, 61}

\bibitem[\protect\citeauthoryear{{Shorlin}, {Wade}, {Donati}, {Landstreet},
  {Petit}, {Sigut}  \& {Strasser}}{{Shorlin} et~al.}{2002}]{Shorlin2002}
{Shorlin} S.~L.~S.,  {Wade} G.~A.,  {Donati} J.-F.,  {Landstreet} J.~D.,
  {Petit} P.,  {Sigut} T.~A.~A.,   {Strasser} S.,  2002, \mn@doi [\aap]
  {10.1051/0004-6361:20021192}, \href
  {http://adsabs.harvard.edu/abs/2002A%26A...392..637S} {392, 637}

\bibitem[\protect\citeauthoryear{{Silaj} \& {Landstreet}}{{Silaj} \&
  {Landstreet}}{2014}]{Silaj2014}
{Silaj} J.,  {Landstreet} J.~D.,  2014, \mn@doi [\aap]
  {10.1051/0004-6361/201321468}, \href
  {http://esoads.eso.org/abs/2014A%26A...566A.132S} {566, A132}

\bibitem[\protect\citeauthoryear{{Smalley} et~al.,}{{Smalley}
  et~al.}{2011}]{2011A&A...535A...3S}
{Smalley} B.,  et~al., 2011, \mn@doi [\aap] {10.1051/0004-6361/201117230},
  \href {http://adsabs.harvard.edu/abs/2011A%26A...535A...3S} {535, A3}

\bibitem[\protect\citeauthoryear{{Smalley} et~al.,}{{Smalley}
  et~al.}{2017}]{2017MNRAS.465.2662S}
{Smalley} B.,  et~al., 2017, \mn@doi [\mnras] {10.1093/mnras/stw2903}, \href
  {http://adsabs.harvard.edu/abs/2017MNRAS.465.2662S} {465, 2662}

\bibitem[\protect\citeauthoryear{{Sokolov}}{{Sokolov}}{1998}]{Sokolov1998}
{Sokolov} N.~A.,  1998, \mn@doi [Astronomy and Astrophysics Supplement Series]
  {10.1051/aas:1998226}, \href
  {https://ui.adsabs.harvard.edu/\#abs/1998A&AS..130..215S} {130, 215}

\bibitem[\protect\citeauthoryear{{Stellingwerf}}{{Stellingwerf}}{1978}]{Stellingwerf1978}
{Stellingwerf} R.~F.,  1978, AJ, 83, 1184

\bibitem[\protect\citeauthoryear{{Stibbs}}{{Stibbs}}{1950}]{Stibbs1950}
{Stibbs} D.~W.~N.,  1950, \mn@doi [\mnras] {10.1093/mnras/110.4.395}, \href
  {http://esoads.eso.org/abs/1950MNRAS.110..395S} {110, 395}

\bibitem[\protect\citeauthoryear{{Tang}, {Chen}, {Chiang}, {Jose}, {Herczeg}
  \& {Goldman}}{{Tang} et~al.}{2018}]{Tang2018}
{Tang} S.-Y.,  {Chen} W.~P.,  {Chiang} P.~S.,  {Jose} J.,  {Herczeg} G.~J.,
  {Goldman} B.,  2018, \mn@doi [\apj] {10.3847/1538-4357/aacb7a}, \href
  {http://adsabs.harvard.edu/abs/2018ApJ...862..106T} {862, 106}

\bibitem[\protect\citeauthoryear{{Totochava} \& {Zhiljaev}}{{Totochava} \&
  {Zhiljaev}}{1981}]{1981AN....302..219T}
{Totochava} A.~G.,  {Zhiljaev} B.~E.,  1981, \mn@doi [Astronomische
  Nachrichten] {10.1002/asna.2103020503}, \href
  {http://esoads.eso.org/abs/1981AN....302..219T} {302, 219}

\bibitem[\protect\citeauthoryear{{Trumpler}}{{Trumpler}}{1938}]{Trumpler38}
{Trumpler} R.~J.,  1938, \mn@doi [Lick Observatory Bulletin]
  {10.5479/ADS/bib/1938LicOB.18.167T}, \href
  {http://adsabs.harvard.edu/abs/1938LicOB..18..167T} {18, 167}

\bibitem[\protect\citeauthoryear{{Ventura} \& {Rodono}}{{Ventura} \&
  {Rodono}}{1993}]{1993MNRAS.263..742V}
{Ventura} R.,  {Rodono} M.,  1993, \mn@doi [\mnras] {10.1093/mnras/263.3.742},
  \href {http://esoads.eso.org/abs/1993MNRAS.263..742V} {263, 742}

\bibitem[\protect\citeauthoryear{{Walker}, {Matthews}, {Kuschnig}  \& {et
  al.}}{{Walker} et~al.}{2003}]{2003PASP..115.1023W}
{Walker} G.,  {Matthews} J.,  {Kuschnig} R.,   {et al.} 2003, \mn@doi [PASP]
  {10.1086/377358}, \href {http://adsabs.harvard.edu/abs/2003PASP..115.1023W}
  {115, 1023}

\bibitem[\protect\citeauthoryear{{Weiss}}{{Weiss}}{1983}]{1983AA...128..152W}
{Weiss} W.~W.,  1983, \aap, \href
  {http://esoads.eso.org/abs/1983A%26A...128..152W} {128, 152}

\bibitem[\protect\citeauthoryear{{Weiss}, {Breger}  \& {Rakosch}}{{Weiss}
  et~al.}{1980}]{1980AA....90...18W}
{Weiss} W.~W.,  {Breger} M.,   {Rakosch} K.~D.,  1980, \aap, \href
  {http://esoads.eso.org/abs/1980A%26A....90...18W} {90, 18}

\bibitem[\protect\citeauthoryear{{Xiong}, {Deng}, {Zhang}  \& {Wang}}{{Xiong}
  et~al.}{2016}]{2016MNRAS.457.3163X}
{Xiong} D.~R.,  {Deng} L.,  {Zhang} C.,   {Wang} K.,  2016, \mn@doi [\mnras]
  {10.1093/mnras/stw047}, \href
  {http://adsabs.harvard.edu/abs/2016MNRAS.457.3163X} {457, 3163}

\bibitem[\protect\citeauthoryear{{Yushchenko} et~al.,}{{Yushchenko}
  et~al.}{2015}]{Yushchenko2015}
{Yushchenko} A.~V.,  et~al., 2015, \mn@doi [\aj] {10.1088/0004-6256/149/2/59},
  \href {https://ui.adsabs.harvard.edu/#abs/2015AJ....149...59Y} {149, 59}

\bibitem[\protect\citeauthoryear{{Zucker} \& {Mazeh}}{{Zucker} \&
  {Mazeh}}{1994}]{Zucker1994}
{Zucker} S.,  {Mazeh} T.,  1994, \mn@doi [\apj] {10.1086/173605}, \href
  {https://ui.adsabs.harvard.edu/#abs/1994ApJ...420..806Z} {420, 806}

\bibitem[\protect\citeauthoryear{{Zverko}}{{Zverko}}{1987}]{1987CoSka..16....7Z}
{Zverko} J.,  1987, Contributions of the Astronomical Observatory Skalnate
  Pleso, \href {http://esoads.eso.org/abs/1987CoSka..16....7Z} {16, 7}

\makeatother
\end{thebibliography}

\bsp	
\label{lastpage}
\end{document}